\documentclass[twocolumn, aps,prl,10pt,showpacs,noeprint,floatfix]{revtex4-1}
\usepackage{bm}
\usepackage{textcomp}
\usepackage{bbold}
\usepackage{hyperref}
\usepackage{url}
\usepackage{amsmath}
\usepackage{amssymb}
\usepackage{amsxtra}
\usepackage{amscd}
\usepackage{amsthm}
\usepackage{amsfonts}
\usepackage{eucal}
\usepackage{latexsym,amsthm}
\usepackage{hhline}
\usepackage{pstricks}
\usepackage{color}
\usepackage{graphicx}

\newcommand{\nc}{\newcommand}
\nc{\on}{\operatorname}
\nc{\wt}{\widetilde}
\nc{\Wick}{{\mathbb :}}
\nc{\R}{{\mathbb R}}

\newcommand{\beq}{\begin{equation}}
\newcommand{\eeq}{\end{equation}}
\newcommand{\bmul}{\begin{multline}}
\newcommand{\emul}{{\end{multline}}}
\newcommand\beqa{\begin{eqnarray}}
\newcommand\eeqa{\end{eqnarray}}
\newcommand\bea{\begin{array}}
\newcommand\eea{\end{array}}
\newcommand\ba{\begin{array}}
\newcommand\ea{\end{array}}
\newcommand{\nn}{\nonumber}

\newcommand{\neqa}{\nonumber\end{eqnarray}}

\newcommand{\eq}[1]{Eq.(\ref{#1})}

\renewcommand{\d}{\partial}

\renewcommand{\L}{{\cal L}}

\renewcommand{\div}{\mathrm{div}}

\nc{\CH}{{\mathcal H}}
\nc{\Db}{{\bar D}}
\nc\comment[1]{}

\nc{\CM}{{\mathcal M}}
\nc{\CN}{{\mathcal N}}

\newcommand{\re}{\relax{\rm I\kern-.18em R}}

\nc{\meV}{{\mathrm{\,meV}}}
\nc{\cG}{{\mathcal G}}

\renewcommand{\bar}{\overline} 

\def\eV{{\mathrm{eV}}}
\nc{\al}{{\alpha}}

\def\eps{{\epsilon}}

\renewcommand\red{{}}
\renewcommand{\)}{\right)}
\renewcommand{\(}{\left(}

\begin{document}
\title{Cooling of chiral heat transport in the  quantum Hall effect graphene}
\author{Sergey Slizovskiy }
\email{On leave of absence from NRC ``Kurchatov Institute'' PNPI, Russia.}
\author{Vladimir Fal'ko}
\affiliation{National Graphene Institute, The University of Manchester, Booth St.E., M13 9PL, Manchester, UK 
}

\begin{abstract}  
In the quantum Hall effect (QHE) regime, heat is carried by electrons in the edge states of Landau levels.
Here, we study cooling of hot electrons propagating along the edge of graphene  at the filling factor $\nu=\pm2$, mediated by acoustic phonons. 
We determine the temperature profile  extended
from a hot spot, where the Hall current is injected into graphene from a metallic contact, taking into account specifics of boundary conditions for lattice displacements in graphene
in a van der Waals heterostructure with an insulating substrate.
Our calculations, performed using generic boundary conditions for Dirac electrons, show that emission of phonons can explain a  
short cooling length observed in graphene-based  QHE devices by Nahm, Hwang and Lee [PRL 110, 226801 (2013)].
\end{abstract}
\maketitle
\begin{figure}[t]
\begin{center}
 \includegraphics[scale=0.42,angle=0]{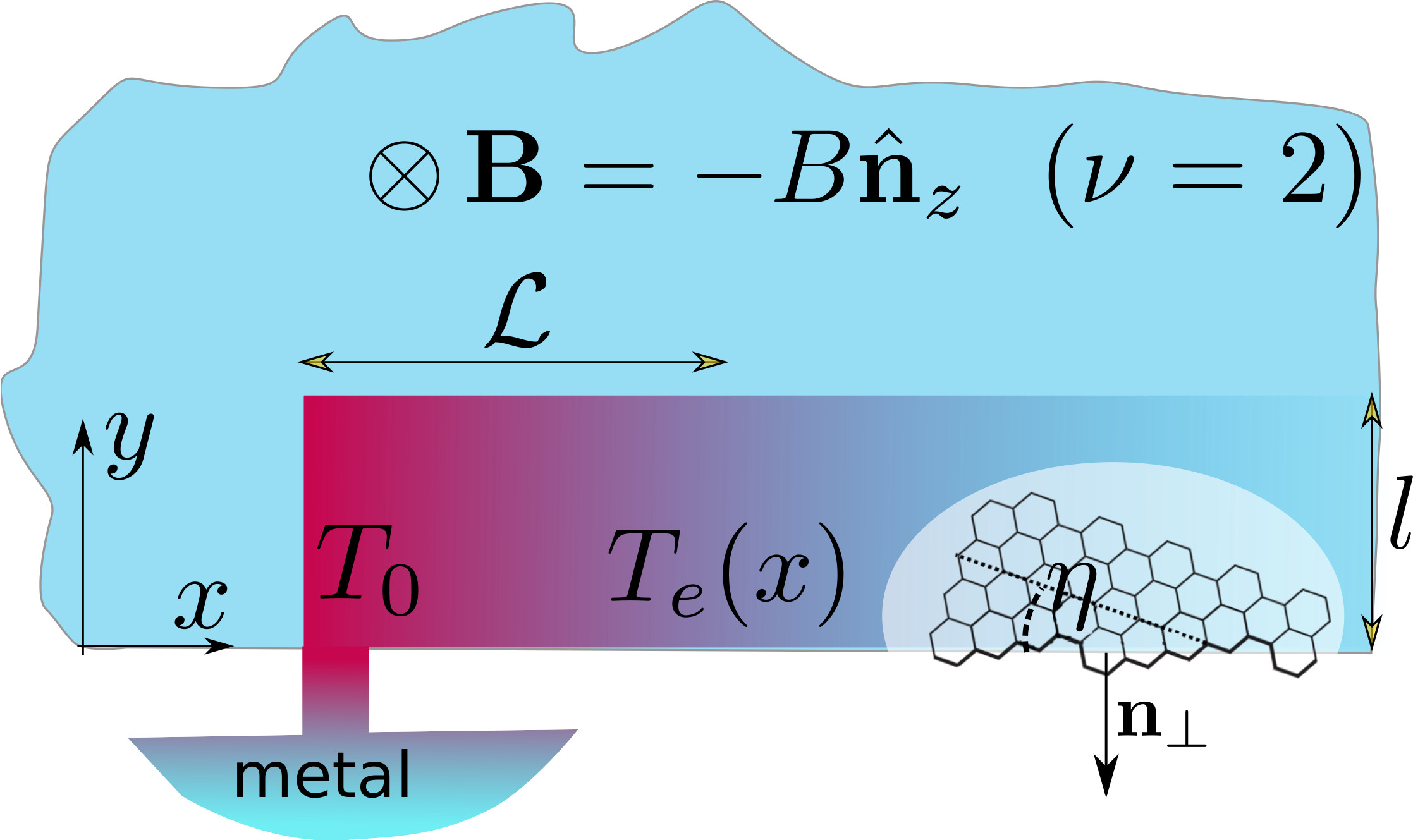}
\end{center}
\vspace{-0.5 cm}
\caption{\label{fig:Illustration} 
Heat injection by a hot spot near metallic contact to graphene in the QHE regime. Edge is oriented at angle $\eta$ from the armchair direction. }
\end{figure}

Graphene offers a promising material platform for the realisation of metrological  resistance 
standard   
based on the phenomenon of Quantum Hall Effect (QHE) \cite{vonKlitzing80,Laughlin81,vonKlitzing86, Sasha1,Sasha2,Rozhko,Sasha3,TransferredGrapheneQHE,CWDGrapheneQHE,LongPlateau16,CWDTransferred}. 
Precision metrology 
requires wide QHE plateux, such as observed at filling factor $\nu=\pm2$ in epitaxial graphene on SiC\cite{Sasha3,LongPlateau16}, and the largest possible breakdown current. 
The breakdown of QHE is triggered by the unavoidable hot spots \cite{vonKlitzing86, VonKlitzing, Komiyama}  formed at the points where current is injected into the 2D electron 
gas from a normal metal contact. 
Hot electrons propagating from the hot spot to the potential contacts used in the QHE measurements
spoil the precision of QHE resistance quantization \cite{EnergyRelaxationGraphene12}.
In the  QHE regime, electron energy spectrum is gapped inside the 2D structure, while the edge states \cite{ButtikerChiral, EdgeHalperin, EdgeThouless1,EdgeThouless2} 
carry non-equilibrium electrons.
The chiral nature\cite{ButtikerChiral, EdgeHalperin} of edge states means  that electrons propagate 
in only one direction, making the heat transport  uni-directional, Fig.\ref{fig:Illustration}, with the direction of the drift reversed upon reversal of magnetic field,
${\bf B} =  -B \hat{\bf n}_z$. Chiral heat transport 
was experimentally observed  both in semiconductor heterostructures \cite{2DEGHeat}  and in graphene \cite{GrapheneExperiment}. 
Here, we propose a theory describing the temperature profile formed by the competition of chiral heat transport and acoustic-phonon-mediated cooling of electrons in the QHE edge states 
in graphene with filling factor $\nu = \pm 2$, for which the gap between LLs is the widest.



To model edge states in graphene, we use Dirac equation for electrons with generic boundary conditions \cite{FalkoBoundary, Akhmerov1, Akhmerov2} that provide zero value for the current
across the edges and satisfy the time-inversion symmetry requirement at ${\bf B} = 0$,
\beq 
\ba{c}
  v  \bm{\sigma} \cdot (-i \hbar {\bf \nabla} + e {\bf A}) \Psi = E \Psi \; ;  \\
  \left[ 1 - ({\bm m} \cdot \bm{\tau}) \otimes ( {\bm n} \cdot \bm{\sigma} )]  \Psi \right|_{y=0}  = 0; \\
  {\bm n} =  \hat{\bf n}_z\, \cos \phi +  [{\hat{\bf n}_z} \times {\bf n_\perp}]\, \sin \phi .
 \ea
 \label{DiracGen} 
\eeq
Here,  $\sigma_i$ and $\tau_i$ are Pauli matrices acting separately on sublattice ($A,B$) and valley ($\pm K$) components of a 4-spinor,  
$\Psi^T = (\Psi_{KA}, \Psi_{KB}, \Psi_{-KB}, -\Psi_{-KA})$, describing
the electron amplitudes on sublattices $A$ and $B$ in the valleys $\pm K$. 
Generic boundary conditions  in \eq{DiracGen} are parameterized by two unit vectors, $\bm m$ 
and 
${\bm n}  \perp {\bf n_\perp}$, where $\bf{n}_{\perp}$ is  normal to
the edge in the 2D plane of graphene, Fig.\ref{fig:Illustration}. 
Both $\bm m$ and $\bm n$ depend on the microscopic features of the edge in a particular sample. In particular, 
deviations of $\bm m$ from 
$\hat{\bf n}_z$ reflect the probability of inter-valley scattering upon specular reflection of an electron arriving at the incidence angle $\theta$ to the edge,
$$P_{K \to -K} = \frac{(\tan \theta)^2}{(\cos \phi)^2 + (\tan \theta)^2} |{\bm m }\times \hat{\bf n}_z|^2. $$ 
Vector $\bm m$ also accounts for the crystallographic 
orientation, $\eta$, of the edge:  for zigzag edge ($\eta = \frac{\pi}{2}$) ${\bm m} = \hat{\bf n}_z$,  
for the armchair edge ($\eta=0$) ${\bm m} = \hat{\bf n}_x$ with $\phi=\pm \frac{\pi}{2}$.
For straight edges, by exploiting valley degeneracy 
in \eq{DiracGen} boundary conditions with different ${\bm m}$'s
can be reduced by a unitary transformation in valley space to the ones with ${\bm m} = \hat{\bf n}_z$, thus, leaving all the sample-specific microscopic features of
the edge incorporated in a single parameter $\phi$. For example,  $\phi=0$ corresponds to an 
idealised nearest-neighbour-hopping model of zigzag edge with equal on-site energies on carbons, which imposes an 
artificial electron-hole symmetry on the electronic spectrum \cite{EdgeStates1, EdgeStates2,EdgeStates3, Abanin2007, Lado}. To compare,
$\phi = \pm \pi/2$ correspond to the 'infinite-mass' boundary 
condition for Dirac fermions \cite{Berry}, due to the sublattice symmetry breaking by a staggered potential, $\Delta \, \sigma_z \otimes \tau_z $, 
with $|\Delta| \to \infty$  at the edge. 
According to Ref. \cite{Akhmerov1}, boundary parameter $\phi$ can be related to  staggered potential $-\Delta \sigma_z \otimes \tau_z $ ␮on  $2N$ rows
at a zigzag edge, as   
$\cos \phi ␪ = \frac{1 + \sinh(\kappa) \sinh(\kappa + 2 N \Delta/t )}{\cosh(\kappa) \cosh(\kappa+ 2 N \Delta/t)}$,
 with $\sinh \kappa ␬ = ␮\Delta/ (2 t)$ and $t$ being the nearest-neighbour hopping in the tight-binding model.
 The boundary parameter $\phi$ also determines a phase shift, $\gamma_\xi = \pi + 2 \, \arctan\frac{\sin \theta \, (\cot(\phi/2))^\xi}{ \xi - \cos \theta \, (\cot(\phi/2))^\xi  }$,
 of a plane wave upon specular reflection of an electron arriving at the incidence angle $\theta$ to the edge at ${\bf B}=0$. 
Note that the armchair edge boundary condition is reduced to $\phi = \pm \pi/2$ (after the abovementioned unitary transformation in the valley space). 
In general, a non-zero $\phi$ determines 
the dispersion of the edge states characteristic for graphene edge at $B=0$, 
\beq \label{SurfaceStates}
E(p) = \xi \hbar v p \sin \phi, 
\eeq
where a valley-specific requirement $\xi p \cos \phi >0$  ($\xi = \pm 1$ for $\pm K$ valleys) guarantees confinement of evanescent states
near the edge, $\Psi_\xi =  \left[\ba{c} \xi \\ \left( \tan \frac{\phi}{2} \right)^\xi \ea \right] e^{- \xi p y \cos \phi+i p x}$.

To analyse the edge state problem (\ref{DiracGen}) for the geomery of the edge  shown in Fig. \ref{fig:Illustration}  ($x/y$ axes chosen  along / 
perpendicular to graphene edge) and an arbitrary choice of $\phi$ and ${\bm m}$,  we employ a unitary transformation, 
 $$\Psi =e^{i \frac{\widehat{{\bm m}\hat{\bf n}}_z}{2} {\bm \tau} \cdot \frac{{\bm m} \times \hat{\bf n}_z}{|{\bm m} \times \hat{\bf n}_z|}} \otimes  e^{-i \eta \frac{\sigma_z}{2}} \Psi', $$
which adjusts  the sublattice composition of spinors in \eq{DiracGen} to the choice of the coordinate
system, simplifying the form of boundary condition by rotating $\bm m$ (to the angle $\widehat{{\bm m}\hat{\bf n}}_z$) to $\hat {\bf n}_z$ axis. 
Then, we use Landau gauge ${\bf A} = (B y, 0)$ for magnetic 
 field ${\bf B} = - B {\bf \hat n}_z$, seeking solutions in the form of 
$\Psi'(x,y) = e^{i p x} \psi(y) $, where $\psi^T=(\psi_{+A},\psi_{+B}, \psi_{-B},-\psi_{-A})$. 
Now, it is convenient to  rewrite the Dirac equation and its boundary conditions in  a 
Schr\"odinger-like form using amplitudes  $\chi_{+1} \equiv \psi_{+A}$ and $\chi_{-1} \equiv \psi_{-B}$ 
on one sublattice only,
\beq 
\ba{c}
\left[- l^2 \d_y^2 + \left(\frac{y}{l}+ l p \right)^2 - 1\right] \chi_{\xi}(y) =  \(\frac{E l}{\hbar v}\)^2 \chi_{\xi}(y),\\ 
\left. \d_y \chi_{\xi}\right|_{y=0} = \left. \left[- l p + \xi \frac{l E }{\hbar v}  \( \tan \frac{\phi}{2}\)^{\xi} \right]\chi_\xi\right|_{y=0},
\label{BoundaryCond} 
\ea
\eeq
where $\xi=\pm 1$,   and $l = \sqrt{\frac{\hbar}{e B}}$ is a magnetic length. 
\begin{figure}
\begin{center}
\includegraphics[scale=0.33]{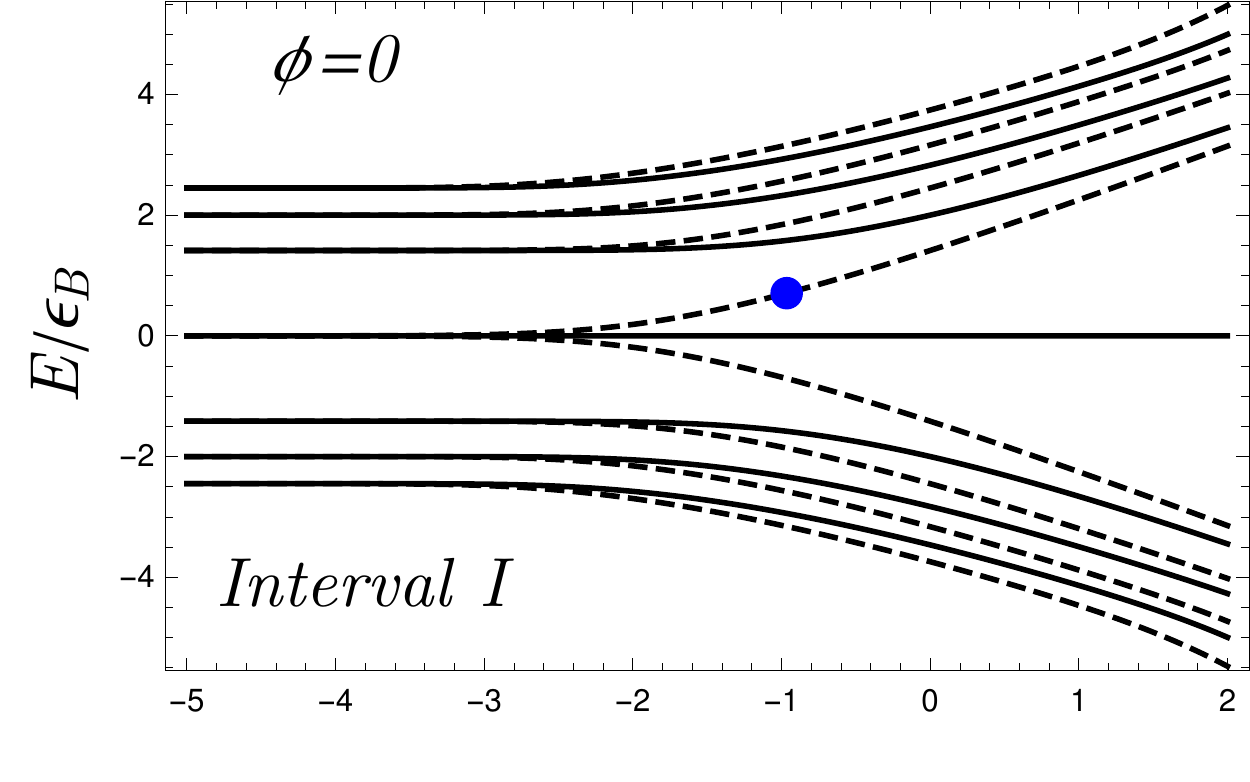}
 \includegraphics[scale=0.32]{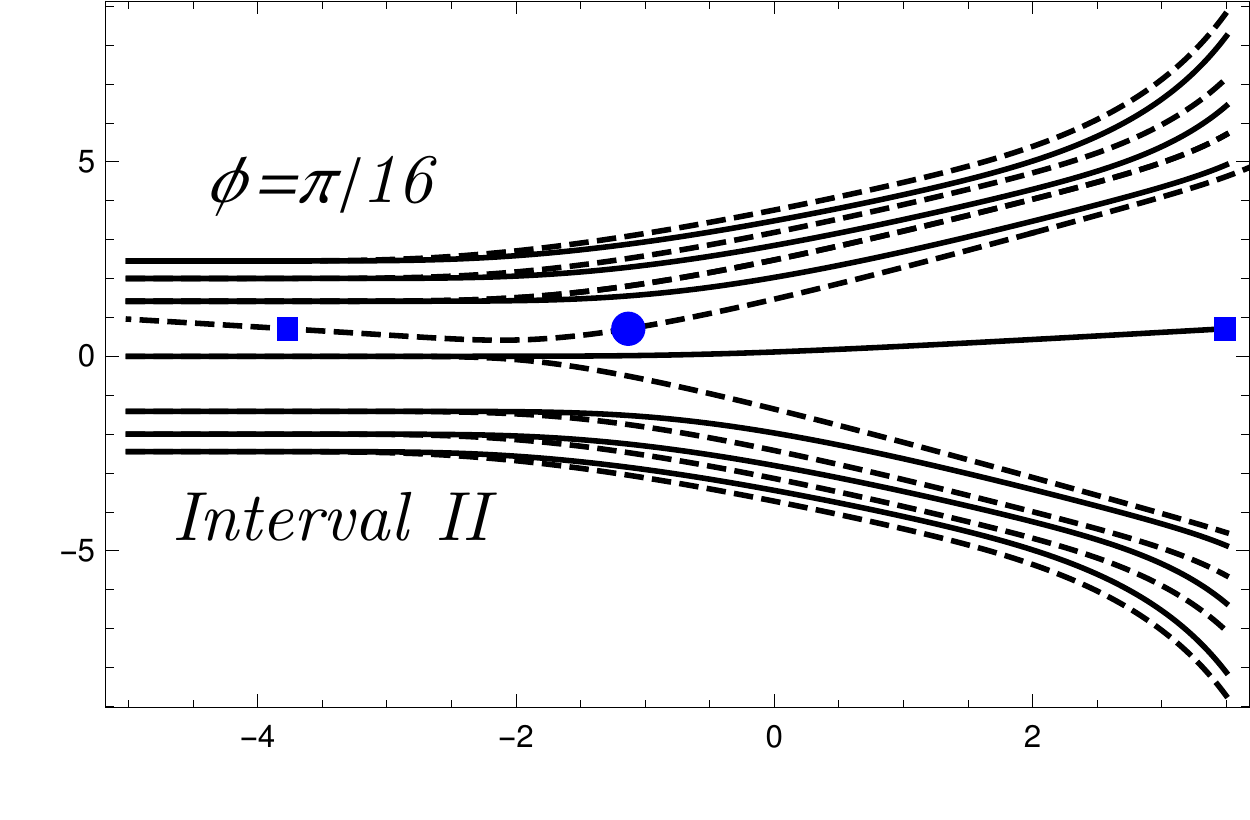}
 \includegraphics[scale=0.33]{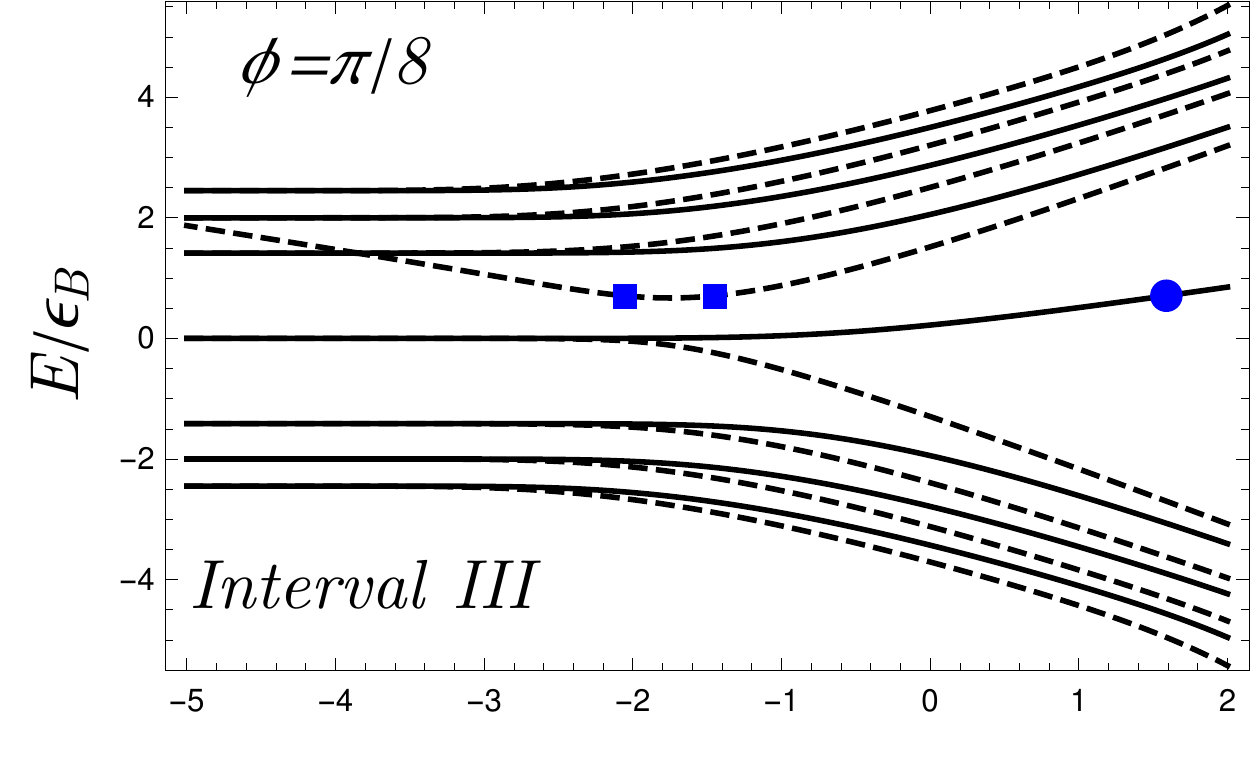}
 \includegraphics[scale=0.32]{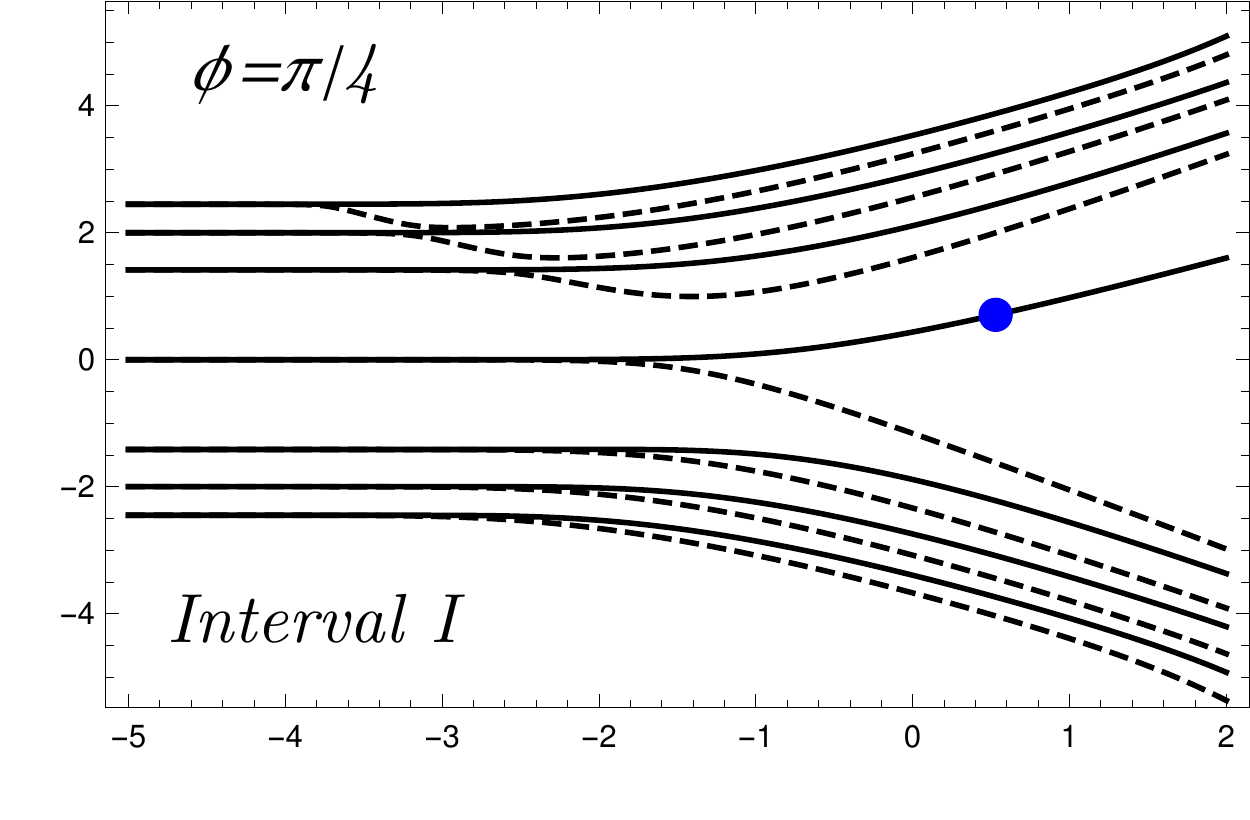}
 \includegraphics[scale=0.33]{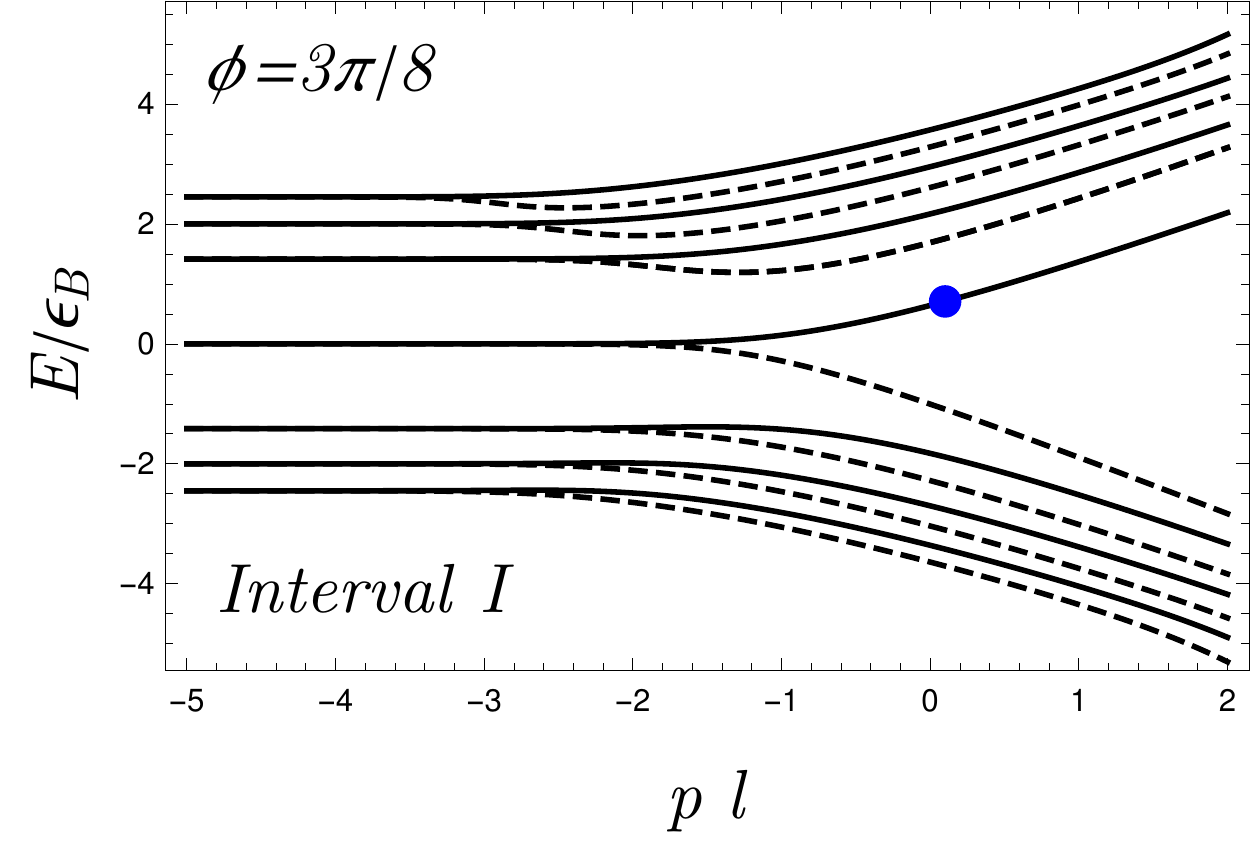}
 \includegraphics[scale=0.32]{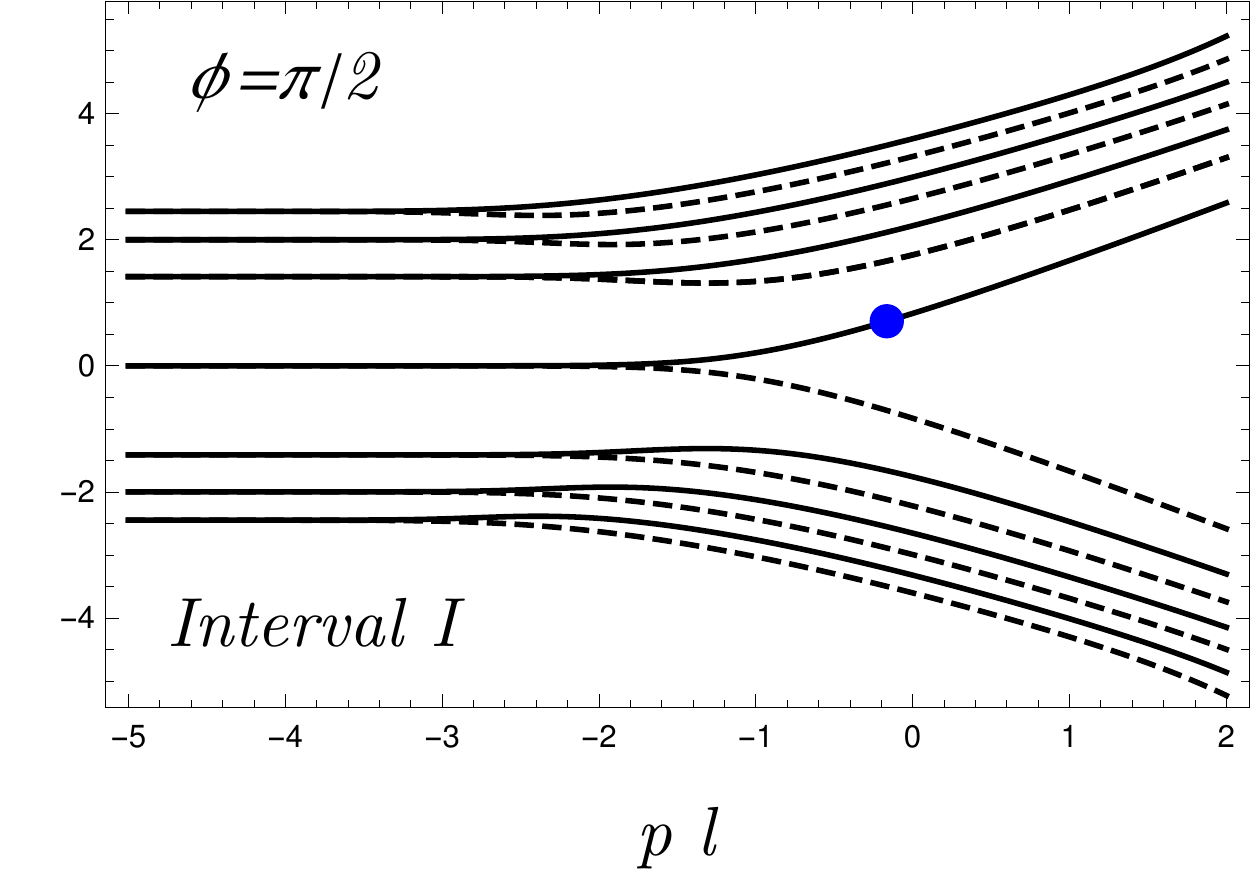}
 \end{center}
\vspace{-0.7 cm}
\caption{\label{fig:BandsPhi}
Bands $E(p)$ of QHE edge states for various boundary conditions, in units of $\eps_B = \hbar v/l$,  $l = \sqrt{\frac{\hbar}{e B}}$.   
Solid/dashed lines correspond to solutions for $\chi_+$ and $\chi_-$ respectively.
 $\nu=2$ edge states at energy in the middle between the 0-th and the 1-st LL ($E_F = \frac{1}{\sqrt{2}}\eps_B $) are indicated by blue circles and square markers, 
 the latter indicating  the additional  edge states present in intervals II and III and expected to get localized if impurities are considered.
Edge states for $\phi \to -\phi$ can be found  by reflection $E(p) \to -E(p)$, and for $\phi \to  \pi + \phi$ by swapping the 
`valleys' $\chi_+ \leftrightarrow \chi_-$.  
}
\end{figure}

Numerical solution of \eq{BoundaryCond} produces spectra $E(p)$ for electrons near the edge shown in Fig.\ref{fig:BandsPhi} for
various values of $\phi \in [0, \frac{\pi}{2}]$.
Spectra 
for inverted, $\phi \to -\phi$, values of the boundary condition parameter can be obtained from those shown in Fig. \ref{fig:BandsPhi} by changing $E(p) \to -E(p)$, and 
spectra for $\phi \to  \pi + \phi$
by swapping  $\chi_+$ and $\chi_-$. 
In Fig.\ref{fig:BandsPhi}, one can recognize  the 2D bulk Landau levels (LLs) at $p<0$, transforming into the edge states near $p=0$ 
The sublattice composition of the corresponding
wave functions is shown in Supplementary Materials (SM) S1. 
Near the edge, the  LLs resonantly mix with the edge states of Dirac electrons in graphene, \eq{SurfaceStates}, whose spectrum and valley 
composition depend on $\phi$. This mixing determines three parametric intervals for the edge states at $E_F = \frac{\hbar v}{\sqrt{2} l}$, which is the optimum choice of
Fermi energy for developing QHE devices\footnote{Moving $E_F$ within the gap introduces no qulitative changes.}:
\renewcommand{\theenumi}{\Roman{enumi})}
\renewcommand{\labelenumi}{\theenumi}
\begin{enumerate}
 \item  $\frac{\pi}{6}<\phi< \pi $ (and $\frac{7 \pi}{6}<\phi< 2 \pi $), where there is one chiral edge state at the Fermi level, whose wave-function 
is a mixture of LLs and the evanescent mode present for $B=0$  for either $\chi_+$ or $\chi_-$ (in only one ``valley''). 
\item  $0<\phi \lesssim \frac{\pi}{8}$ (and $\pi <\phi \lesssim \frac{9 \pi}{8}$), where there are three edge modes.
Two of these modes (one for $\chi_+$ and the other for $\chi_-$) are the counterpropagating evanescent modes with $\psi \propto e^{-\frac{y}{\sqrt{2} l} |\cot \phi|}$
confined at a short distance, $\lambda_{\phi} \sim \sqrt{2} l \tan \phi$, near the edge.  Due to a strong confinement, $\lambda_{\phi} \xrightarrow[{\phi \to 0}]{} 0$, 
these modes have suppressed  scattering into the LL edge state.  
Note that, for $\phi \to 0$, a stronger confinement of evanescent modes would suggest a stronger intervalley scattering on atomically sharp edge disorder, 
localizing the counterpropagating evanescent modes. 
\item $\frac{\pi}{8} \lesssim \phi \lesssim \frac{\pi}{6}$ (and $\frac{9 \pi}{8} \lesssim \phi \lesssim \frac{7\pi}{6}$ ), 
characterized by two additional  counterpropagating modes $\chi_-$($\chi_+$) with strongly overlapping wave-functions  and one chiral mode $\chi_+$($\chi_-$). In this case, ``intra-valley'' scattering would lead to localization of pairs 
of $\chi_-$($\chi_+$) edge states.
\end{enumerate}

As to the phonons, the feature of graphene coupled by van der Waals forces to the underlying substrate with a non-commensurate lattice   is that 
vibrational properties in graphene lattice are only weakly perturbed by the substrate\cite{LowFriction}.
Hence, a graphene flake can be described as a membrane  with free  edges,  where vanishing of stress tensor components in the 
direction ${\bf n}_\perp$ results in a 
zero deformation potential at the edge.
This should be contrasted  to graphene with clamped edges, where the displacement vector field of graphene is set to zero at the edge. 
In principle, both the longitudinal (LA, with displacement field ${\bf u} \parallel {\bf q}$) and  
the transverse (TA, with ${\bf u} \perp {\bf q}$) acoustic phonons in graphene could be emitted by hot electrons in the edge states. Those are
coupled to electrons via a potential, 
$$
 V = g \, \div {\bf u} + g' {\bm \sigma}\cdot {\bf w} , \ \ {\bf w} = (\d_x u_x - \d_y u_y,  - 2 \d_y u_x ), 
$$
with \cite{Coupling1,Coupling2,Coupling3,KaneMele97,Review2009, PhononCoupling} $g \approx 20\,\eV$   and  $g' \approx 2\, \eV$ (since graphene is in the incompressible QHE plateau state, 
we use the unscreened value of the deformation potential coupling). 
As $g' \ll g$, we base our analysis on emission of LA phonons provided by deformation potential with the constant $g$. 
 Note that flexural phonons couple quadratically to electrons, allowing only for two-phonon processes that are negligible at low lattice temperatures \cite{KatsnelsonGeim}. 
Also, we find 
 [Table \ref{tab:parameters} in Supplementary Materials (SM)]  that 
the edge state velocities, $v_e=\left. \frac{\d E(p)}{\d p}\right|_{E=E_F}$,   are much
higher than the speed of longitudinal sound in graphene \cite{GrapheneSound}, $s\approx 2.1 \cdot 10^4 \rm m/s $; therefore,  phonons 
are emitted with the wave vectors almost perpendicular to the edge, ${\bf q} = - q {\bf n}_{\perp}$.  
This feature of electron-phonon (e-ph) interaction  in the edge states simplifies the description of their cooling kinetics.
In particular, for the samples where parameter $\phi$ 
belongs to Interval I, where there is only one edge mode at $E_F = \frac{\hbar v}{\sqrt{2} l}$ for $\nu=2$, 
the e-ph interaction has the form,
\beqa \label{Heph}
 & H_{\text{e-ph}} = -g  \int \frac{d^2 {\bf q}\, d p}{(2 \pi)^3 } \sqrt{\frac{\hbar |q|}{2 \rho  s}} ( 
b_{\bf q} + b_{\bf-q}^\dag ) A(q_y) a^\dag_{p+q_x} a_{p},  
\eeqa
where  $b_{\bf q}^\dag$ ($b_{\bf q}$) and $a_{p}^\dag\,  (a_{p})$  are the creation (annihilation) operators of LA phonons and of edge state electrons;
$\rho \approx 7.6 \cdot 10^{-7} \,{\rm kg/m^2}$ is graphene mass density.
Here, the form factor,
\beq \label{FormFactorDef}
 A(q)= \int_0^\infty dy \sqrt{2} \sin(q y) |\psi(y)|^2, \nn
\eeq
takes into account the form of the edge state wave function and the phonon displacement field near the edge: for the free edge the phonon displacement mode behaves as $u_y \sim \cos(q y)$, 
leading to a deformation potential $V \sim q \sin(q y)$.
For the states marked by dots in Fig. \ref{fig:BandsPhi}, the form factor $A(q)$ has the following characteristic asymptotics,
\beq 
A(q)\approx \left\{ \ba{ccc}  \sqrt{2} r_0 (q l )^{-1}, & q \gg l^{-1}, & r_0 =l | \psi(0)|^2 ; \\
                              \sqrt{2} \alpha_0 q l, &  q \ll l^{-1}, &  \alpha_0 = \int_0^\infty \frac{y}{l} | \psi(y) |^2 dy ; \ea \right.    
\label{Aasympt} \nn
\eeq
values of parameters $r_0$ and $\alpha_0$  
are listed in  Table I of SM.   

 

Edge state electrons cool down  while drifting along the edge. 
Assuming that, locally, their equilibration  due to Coulomb
interaction is fast, electrons' distribution can be
described using a Fermi function with a local temperature $T_e(x)$.  We note that equilibration of edge channels has been shown to be non-thermal at longer time scales 
\cite{Kovrizhin} due to integrability,  but at the comparable time scales
the phonon emission  becomes important, breaking the integrability.  We also find that for a slow thermalization of edge state electrons, their distribution
formed in the course of cooling by phonon emission is almost indistinguishable from thermal, see  SM S2. 
The cooling power (per unit length) provided by the phonon emission by the edge state electrons is
\beq \label{MainFormula3}
 W_{I}=\frac{g^2 T_*^4 }{\hbar^4 \rho v_e^2 s^3} f_{A}\(\frac{T_e}{T_*},\frac{T}{T_*}\).   
\eeq
Here,  $T$ is graphene lattice temperature,  $T_e$ is the temperature of electrons in the edge state, and a factor, 
\beqa \label{MainFormula4} \label{MainFormula01}
  &f_A(\tau, \frac{T}{T_*}) = \frac{\tau^4}{4 \pi^2} \int\limits_0^\infty d\eps \, \eps^3  \left[ A\(\frac{\eps \tau}{l}\)\right]^2 \times  
 \hspace{1cm} \nn \\ &\times\left[ n_B(\eps)  \(1+n_B\(\eps \tau 
\frac{T_*}{T}\)\) - (n_B(\eps)+1) n_B\(\eps \tau \frac{T_*}{T}\)\right]   \nn\\
& \hspace{1.5cm} \xrightarrow[{T\to 0}]{} \frac{\tau^4}{4 \pi^2} \int\limits_0^\infty d\eps \, \eps^3  \left| A\(\frac{\eps \tau}{l}\)\right|^2  n_B(\eps),
\nn
\eeqa
 accounts for all phonons emitted at various energies $\hbar \omega = \eps T_e$, 
with $n_B(\eps) = \left[e^{\eps} -1\right]^{-1}$.
 The temperature scale,
\beq
T_* =   s \sqrt{\hbar e B} \approx 6.5 \sqrt{B{\rm [Tesla]}} \, 
\rm K ,
\eeq 
is determined by the energy of phonons with $q \sim 1/l$.  
It marks a crossover between the two regimes: the high-temperature regime, 
$
T_e \gg T_*
$, where wavelengths of typical emitted phonons are smaller than $l$, $q l > 1$,  and a low-temperature regime,  $T_e \ll T_*$,
where $q l<1$.
Then, for $T \ll T_e$, we find  
\beq \nn
f_A\left(\frac{T_e}{T_*}, \frac{T}{T_*}=0 \right) \approx  \left \{ \ba{cc}  
\frac{r_0^2}{12} \(\frac{T_e}{T_*}\)^2   ,  & T_e \gg T_*; \\
 \frac{4 \pi^4 \alpha_0^2}{63} \(\frac{T_e}{T_*}\)^6, & T_e \ll T_* \ea \right. .
\eeq 
For Interval II, {\red the slow counter-propagating modes (seen in Fig. 2) provide additional cooling power. Since the spatial scale of their wave-functions 
is reduced to $l \tan \phi$, and their velocity is reduced to $v |\sin \phi|$, their cooling power is described by the same \eq{MainFormula3}, but with increased $T_*  \to T_* |\cot \phi|$. 
This leads to the extended range of low-temperature asymptotics, giving a small correction to cooling power }
\beq \label{correctedLocalized}
 W = W_I + 
 \frac{0.62 g^2 T_e^6}{\hbar^4 \rho v^2 s^3 
 T_*^2} {\red , \  T_e  <  T_* \cot \phi },
\eeq 
{\red which rapidly grows to a dominant term at $T_e > T_* \cot \phi$. In  the latter case,  the  equilibration  between the edge channels shortens the cooling length.}
For Interval III, \eq{correctedLocalized} can only give the lowest bound for the cooling efficiency, as phonons can be emitted by both intra- and inter-edge-state transitions. 


All this leads to the equation for the temperature profile $T_e(x)$ along the edge,
\beq
 C \frac{\d T_e}{\d t} + C_e v_e  \frac{\d T_e}{\d x} = - W,  \nn
\eeq
where $C_e = \frac{\pi T_e}{6 \hbar v_e}$ is specific heat of a 1D Fermi gas, $v_e$ is velocity of the fastest propagating edge channel,  
and $C \geq C_e$ is a full specific heat including the
contributions from localized states at the edges.  
{\red The validity of this equation for Intervals II and III is provided by cancellation of $C_e v_e$ contributions from the counter-propagating edge states and assuming that temperature 
equilibration between the edge channels is fast. If inter-channel equilibration is slow, then the effect of additional edge channels can be neglected.}
In the steady state, $\d_t T=0$, this reduces to
\beq  \label{DistanceDecay}
 \frac{\d T_e}{\d x} = \frac{-W}{C_e v_e} .
\eeq 
Also, for comparing our results to the experimental
data from Ref.\onlinecite{GrapheneExperiment}, it is convenient to characterize  
cooling by a differential cooling length, $\L(T_e)$, defined as,
\beq  \label{DistanceDecay1}
 \L^{-1} \equiv \left | \frac{\d \log T_e}{\d x} \right |   = \frac{W}{C_e v_e T_e}.
\eeq 

\begin{figure}[t]
 \includegraphics[scale=0.85]{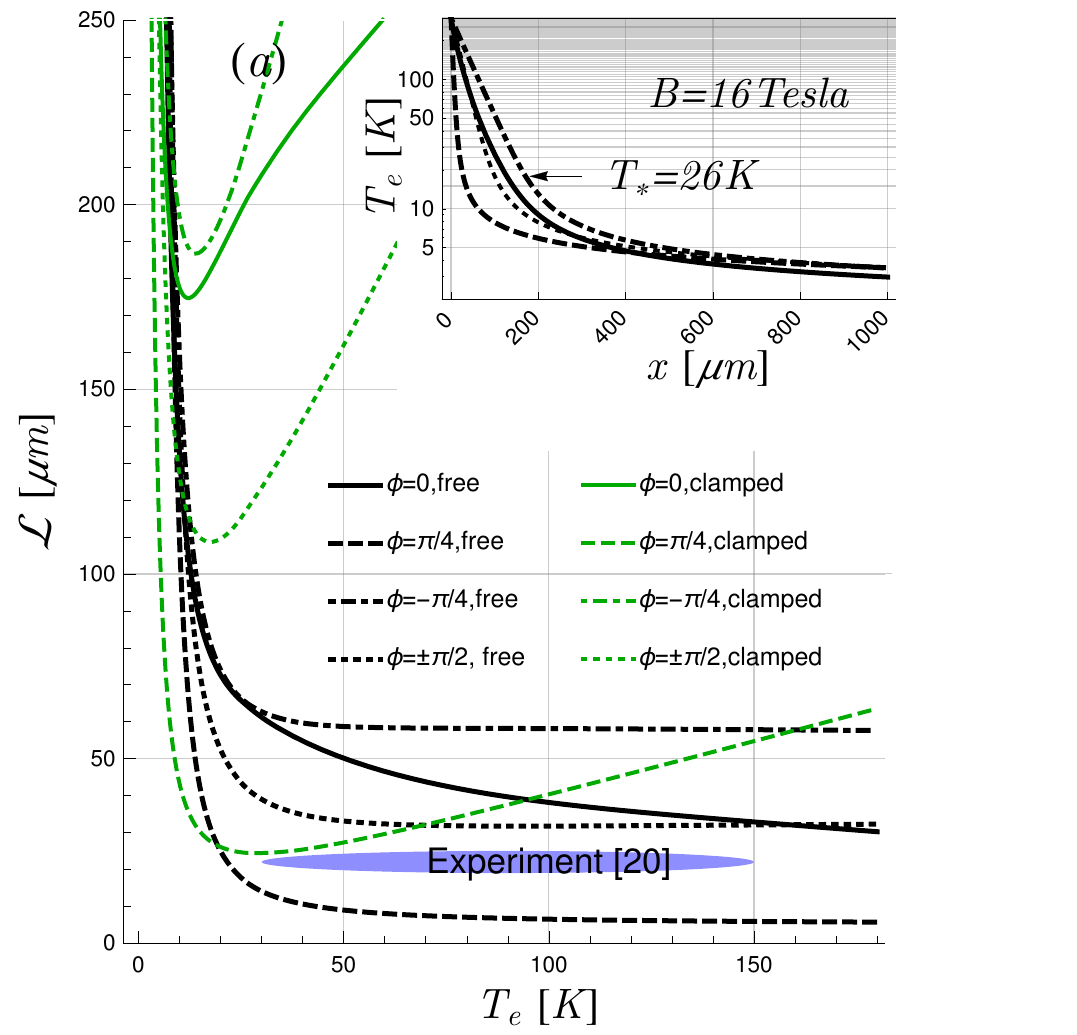}\\
\hspace*{0.3mm} \includegraphics[scale=0.83]{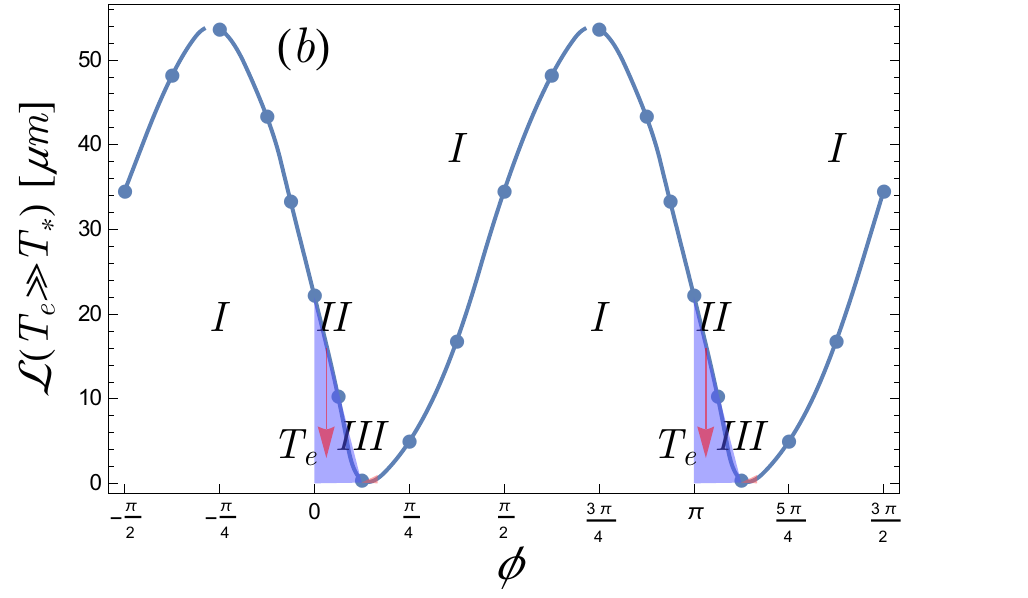}
\vspace*{-0.7 cm}
\caption{\label{fig:Tdecay} \label{fig:CoolingNu6} \label{fig:Lplot}
(a) Differential cooling length (black lines)  computed for graphene with $\phi=0,\ \pm \pi/4$, $\nu=2$ QHE,  
 and $B=16$ Tesla.  
Blue blob shows the experimental value $\L_{exp} \approx 22 \rm \mu m$ measured by Nahm, Hwang and Lee \cite{GrapheneExperiment}.
For comparison, green lines show $\L(T_e)$ for a flake with clamped edges (see SM S3).
Inset: Electron temperature profile, $T_e(x)$, along the edge.
(b) 
High-temperature cooling length, $\L(T_e \gg T)$, for $\nu=2$ QHE edge state at $B=16$ T, plotted  as a  function of boundary parameter $\phi$. 
Triangle-shaped areas indicate
the drop in  $\L(T_e)$ when $T_e$  is above $T_* \cot \phi$, caused by the increasing phonon emission from the pairs of additional counter-propagating  edge states indicated in Fig. 2. 
Results for $\nu = -2 $ are obtained by the reversal $\phi \to -\phi$.
}
\end{figure}

Solving \eq{DistanceDecay}, we find exponentially fast temperature decay at high  temperatures, with   $\gamma =  \frac{g^2 r_0^2 e B}{2 \pi \hbar^2 \rho v_e^2 s}$,
and a slow power-law low-temperature tail:
\beqa \label{HighTCooling}
  &T_e(x) \approx \left \{ \ba{cc} T_0\,  e^{-\gamma x},  &  \! \! T_e > T_* ;\\
  \frac{0.24 \sqrt{r_0/\alpha_0}\, T_*}{\sqrt[4]{\gamma x -  \ln \max\left[1,\frac{T_0}{T_*}\right]  + 0.003 \frac{r_0^2}{\alpha_0^2} \max\left[1,\frac{T_*^4}{T_0^4}\right]}},&  \! \!T_e < T_*.  
  \ea \right. 
    \label{xiResult} \nn
  \eeqa
The crossover between the two asymptotical regimes, from the fast decay at $T> T_*$ to the long tail at $T<T_*$,  is illustrated 
in the inset of Fig.\ref{fig:Tdecay}(a), 
and the values of high-temperature cooling length, 
\beq \label{DecayLength}
 \L(T_e \gg T_*) \approx \frac{2 \pi \hbar^2 \rho v_e^2 s}{g^2 r_0^2 e B},
\eeq
are plotted in Fig. \ref{fig:Lplot}(b). 
It is interesting to note that 
reversing magnetic field does not change the values of the cooling length $\L$, which should be, now, applied to the 
hot electrons drifting in the opposite direction. However, changing $\nu=2$ to $\nu = -2$ by changing  polarity of doping, would be equivalent
to swapping $\phi \to -\phi$ in Fig.3(b), resulting in a different cooling length $\L$ for n- and p-doping  of the same sample when $\phi \neq 0, \pm \frac{\pi}{2}$.

In the recent experiments  on the chiral heat transport along the  QHE edges  
in graphene \cite{GrapheneExperiment},  
 the temperature decay length $\L$ has been measured for the filling factor $\nu=-2$ at $B=16$ T,
resulting in  $\L_{exp} \approx 22 \rm \mu m$.  
Estimates, based on the theory of phonon-assisted scattering  
in semiconducting heterostructures\cite{MartinFeng90} have led the authors of Ref. \onlinecite{GrapheneExperiment} to refute the acoustic phonon cooling mechanism, 
leaving their results
unexplained
\footnote{Here, we neglect the evaporational cooling by emission of hot electrons into the bulk Landau levels, as it is 
suppressed by the Boltzmann factor $e^{-\hbar v/(2 l k_B T_e)} $. For $B=16$ T  $\hbar v/(2 l k_B) = 840 $ K,    
making this mechanism redundant for $T \lesssim 150$ K. }.
Here we rebuff that conclusion.
The values of $\L(T_e)$, computed using \eq{DistanceDecay1} and plotted in Fig.\ref{fig:Tdecay}, show that energy relaxation due to emission 
of LA phonons in graphene with free 
boundary conditions for lattice vibrations does deliver sufficiently 
short  cooling lengths in QHE edge state electrons to explain the value measured by  Nam, Hwang, and Lee \cite{GrapheneExperiment}.  

To summarize, the theory of cooling of hot electrons in the $\nu = \pm 2$ QHE edge states in 
graphene  presented
in this letter shows that for a mechanically free edge of graphene in devices where a flake is bound to the substrate by van der Waals forces, the  temperature profile
extended from a hot spot near a current contact   decays at the length scale $\L \approx \frac{2 \pi \hbar^2 \rho v_e^2 s}{g^2 r_0^2 e B}$  up to where
$T_e(x)$ drops to the value  $T_*  =   s \sqrt{\hbar e B} \approx 6.5 \sqrt{B{\rm [Tesla]}} $. After that (at longer distances), $T_e(x)$ has a long power-law tail 
$T_e(x) \sim (x + x_0)^{-1/4}$.  
The proposed theory explains the earlier measured values of temperature decay 
length. It also shows that the cooling length 
$\L$ would be strongly modified by clamping graphene edge, and it predicts an electron-hole asymmetry of the 
cooling length, which can be tested by changing the polarity 
of doping, from $\nu=+2$ to $\nu = -2$, in the same sample.

We acknowledge useful discussions with K. von Klitzing, S.~Rozhko, A.~Tzalenchuk, J.T.~Janssen, R.~Nicholas, F.~Essler, J.~Wallbank.  
This work is supported by InnovateUK grant and the European Graphene Flagship project.

\clearpage
\begin{widetext}
\begin{center}
\textbf{\large Supplemental Materials:\\ Cooling of chiral heat transport in the  quantum Hall effect graphene} \\ 
 Sergey Slizovskiy,  Vladimir Fal'ko \\
{\it \small National Graphene Institute, The University of Manchester, Booth St.E., M13 9PL, Manchester, UK}
\end{center}
\end{widetext}
\setcounter{equation}{0}
\setcounter{figure}{0}
\setcounter{table}{0}
\setcounter{page}{1}
\makeatletter
\renewcommand{\theequation}{S\arabic{equation}}
\renewcommand{\thefigure}{S\arabic{figure}}
\renewcommand{\bibnumfmt}[1]{[S#1]}
\renewcommand{\citenumfont}[1]{S#1}
\renewcommand{\thesection}[1]{\arabic{section}}

\begin{table*}[!]
 \begin{tabular}{c | c c c c c c c c c c c c c}
  $\phi$ & $-\frac{3 \pi}{8}$& $-\frac{\pi}{4}$ &  $-\frac{\pi}{8}$ & $-\frac{\pi}{16}$ &  $0$ & $\frac{\pi}{16}$ & $\frac{\pi}{8}[1]$& $\frac{\pi}{8}[2]$ & $ \frac{\pi}{8} [3]$ & $\frac{\pi}{4}$ &$\frac{3 \pi}{8} $ & $\pm\frac{\pi}{2}$ \\
  \hline
  $v_e/v$    & 0.77 & 0.78 & 0.75 & 0.71 & 0.65 & 0.53 & -0.23 & 0.24 & 0.34 &0.55 & 0.67 & 0.74 \\
  $r_0$      & 0.74 & 0.71 & 0.76 & 0.82 & 0.92 & 1.1  & 2.9 &1.57 & 3.47 &   1.65 & 1.09 & 0.84\\
  $\alpha_0$ & 0.77 & 0.85 & 0.91 & 0.94 & 0.96 & 0.97 & 0.55& 0.89 &0.26 &   0.44 & 0.58 & 0.69\\
  $r_2$      & 0.03 & 0.02 & 0.02 & 0.02 & 0.04 & 0.05 & 0.3 & 0.12 & 0.4  &  0.15 & 0.06 & 0.03
    \end{tabular}
 \caption{\label{tab:parameters} Parameters of edge states for different examples of boundary conditions. For $\phi = \frac{\pi}{8}$,
 parameters for the three edge states are presented.}
\end{table*}
\section{S1. Wave functions and phonon form-factors for edge states in graphene \label{sec:APlots}}
\begin{figure*}
$\ba{ccc}
\includegraphics[scale=0.5]{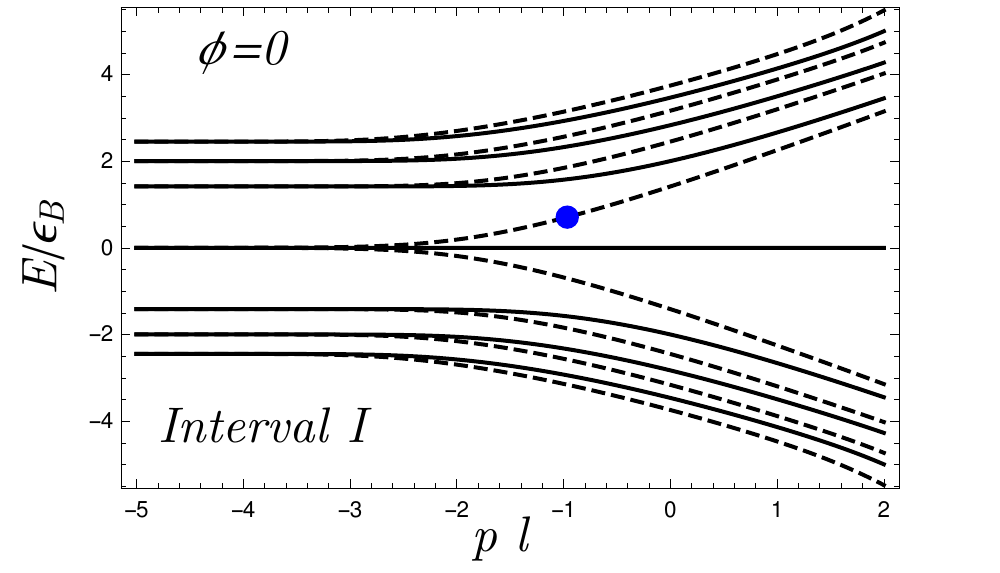} & \includegraphics[scale=0.5]{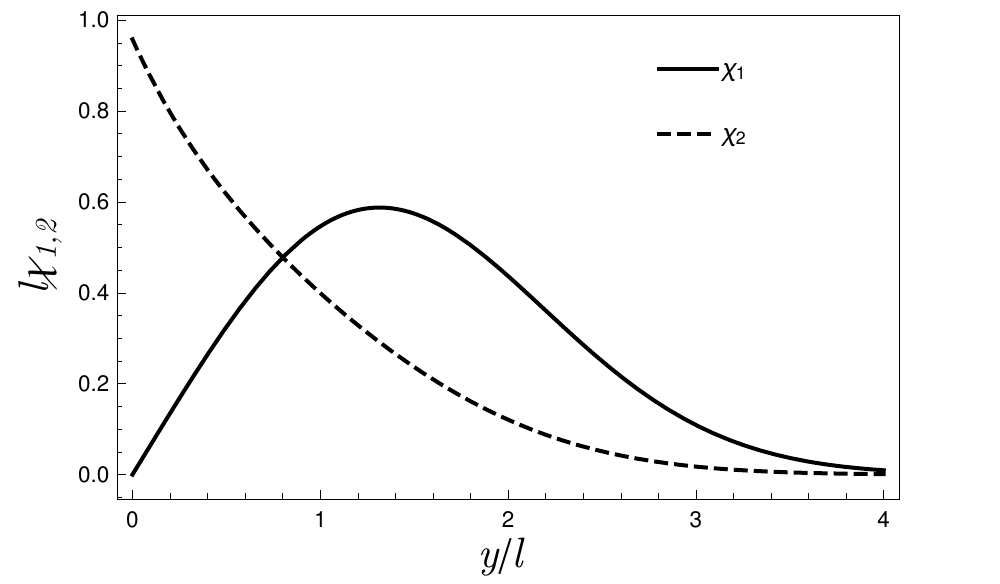} & \includegraphics[scale=0.5]{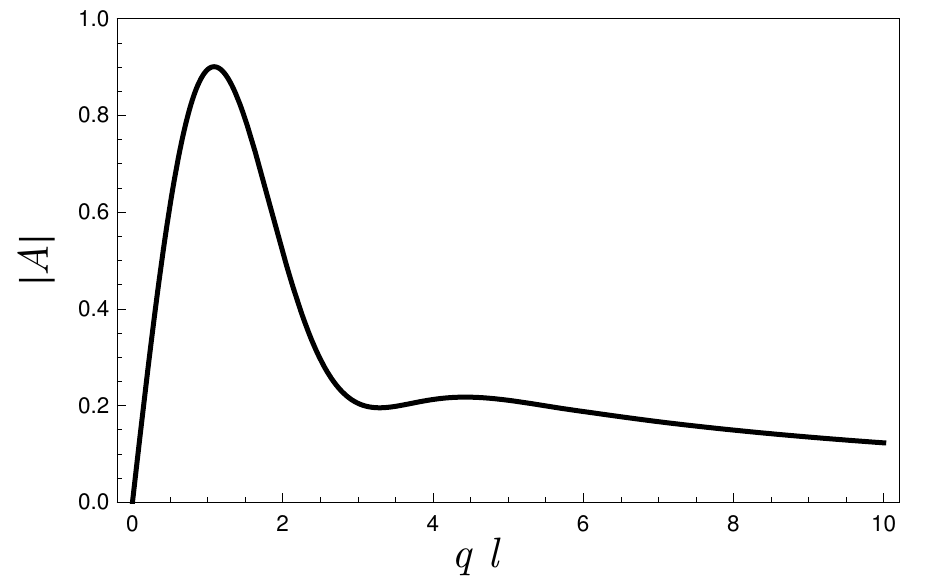}\\
\includegraphics[scale=0.5]{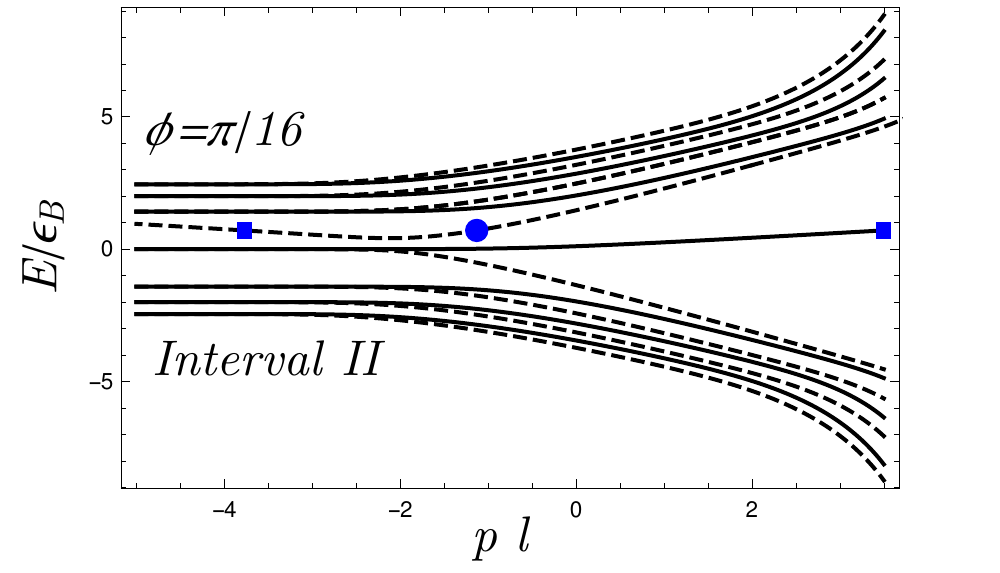} & \includegraphics[scale=0.5]{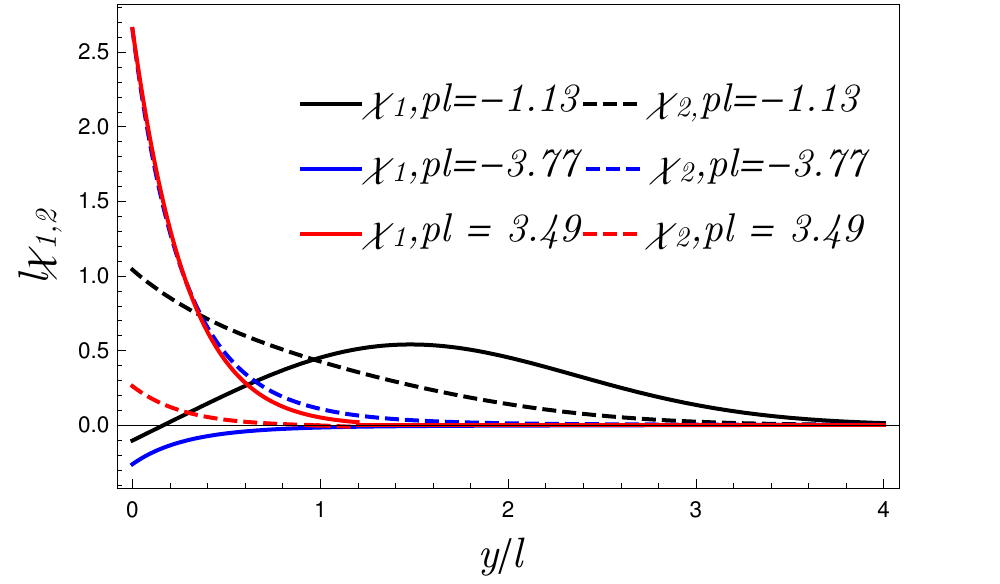} & \includegraphics[scale=0.5]{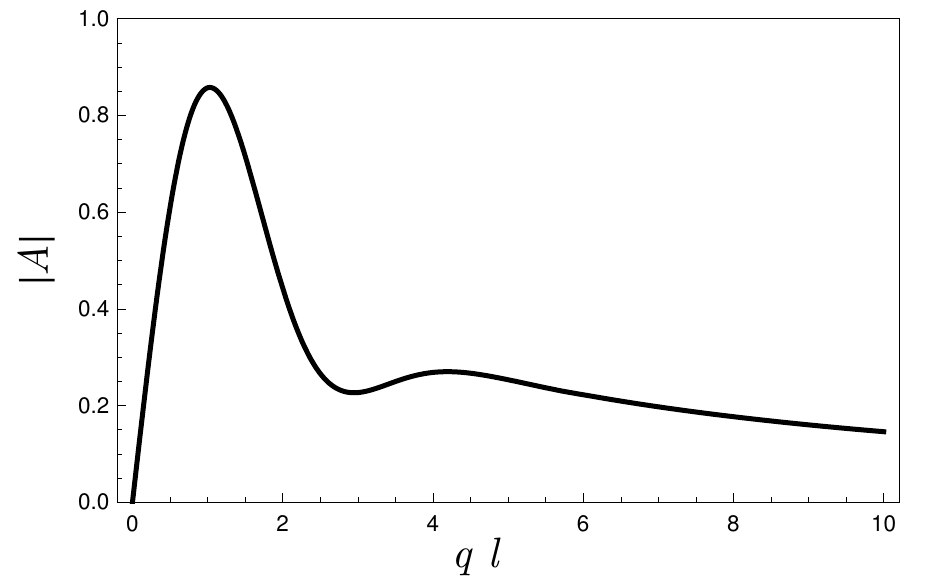}\\
\includegraphics[scale=0.5]{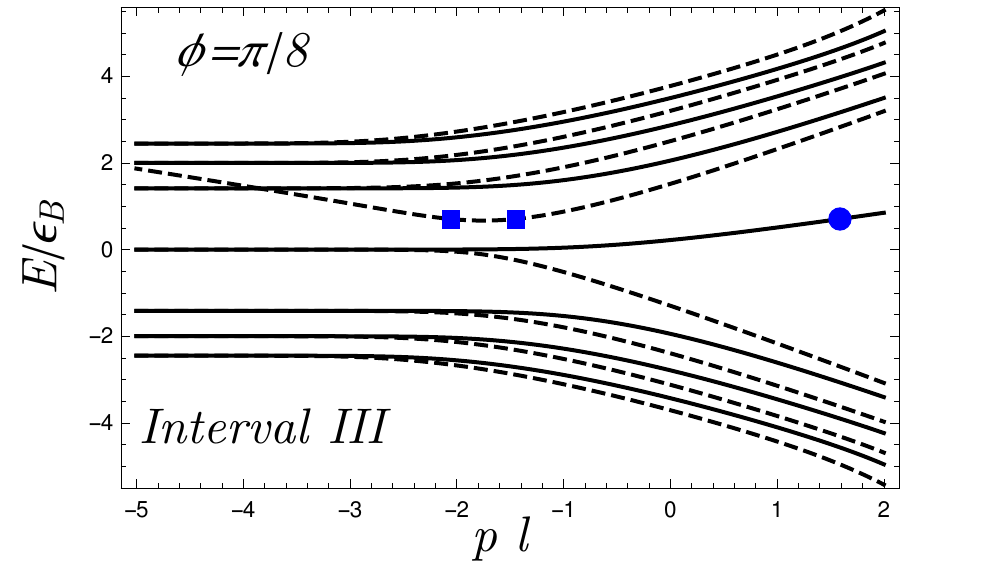} & \includegraphics[scale=0.5]{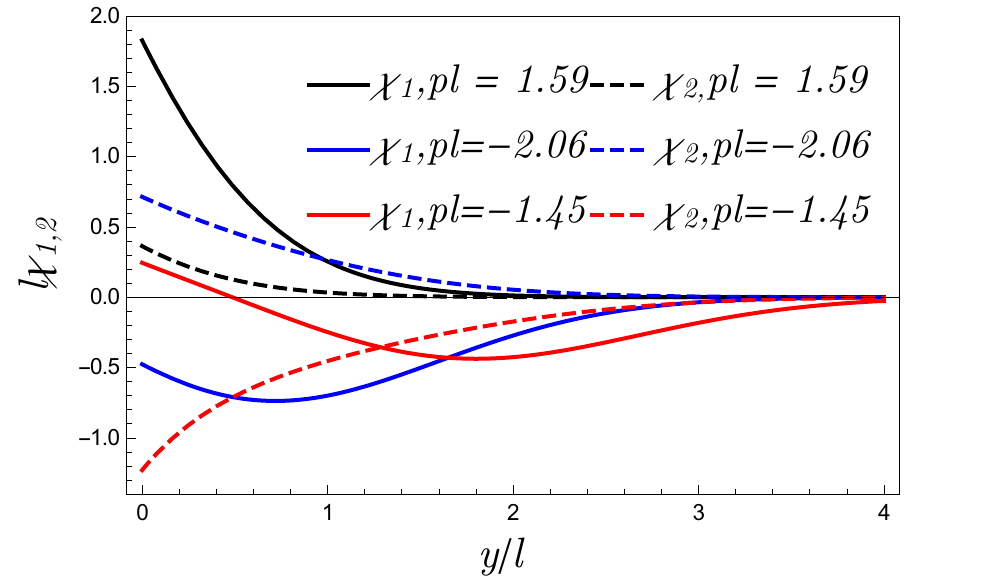} & \includegraphics[scale=0.5]{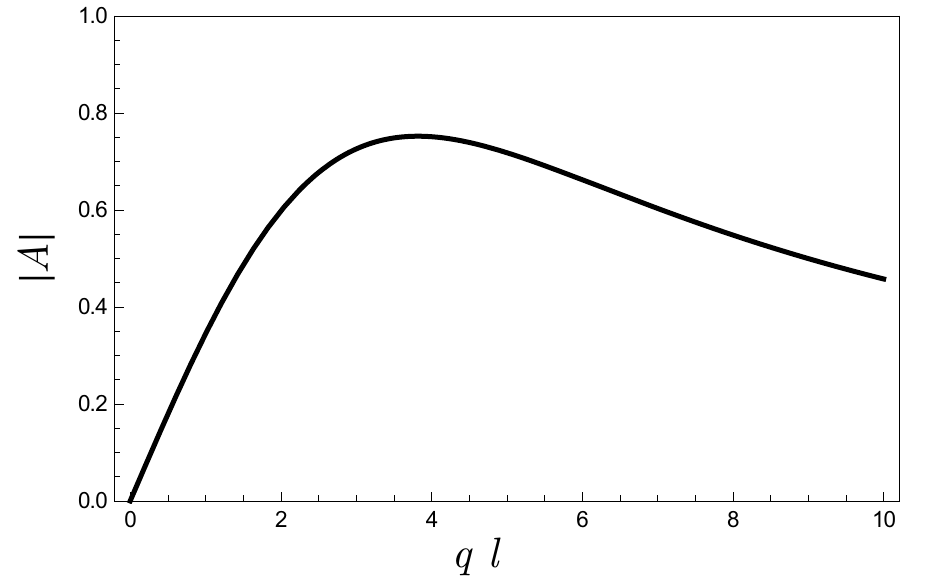}\\
\includegraphics[scale=0.5]{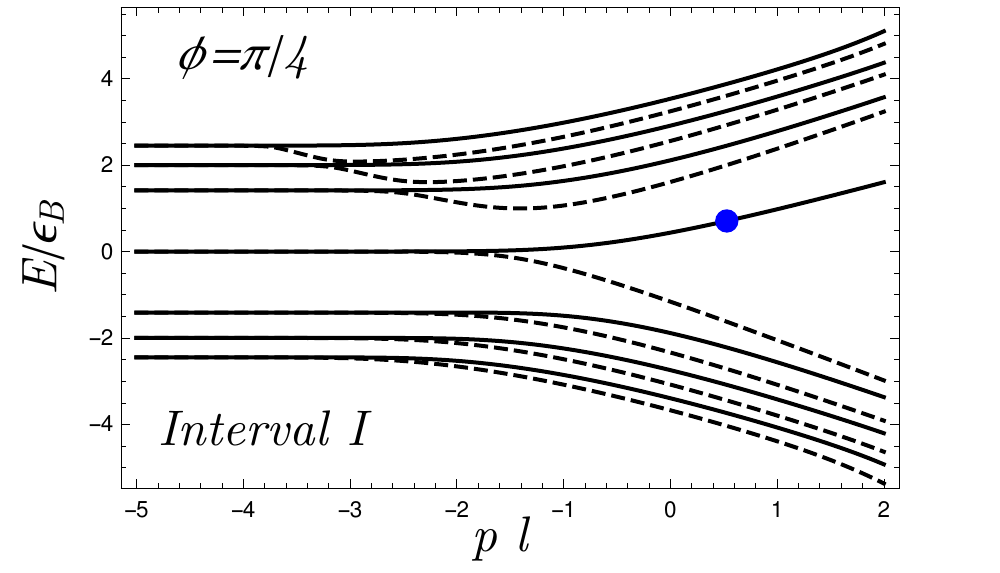} & \includegraphics[scale=0.5]{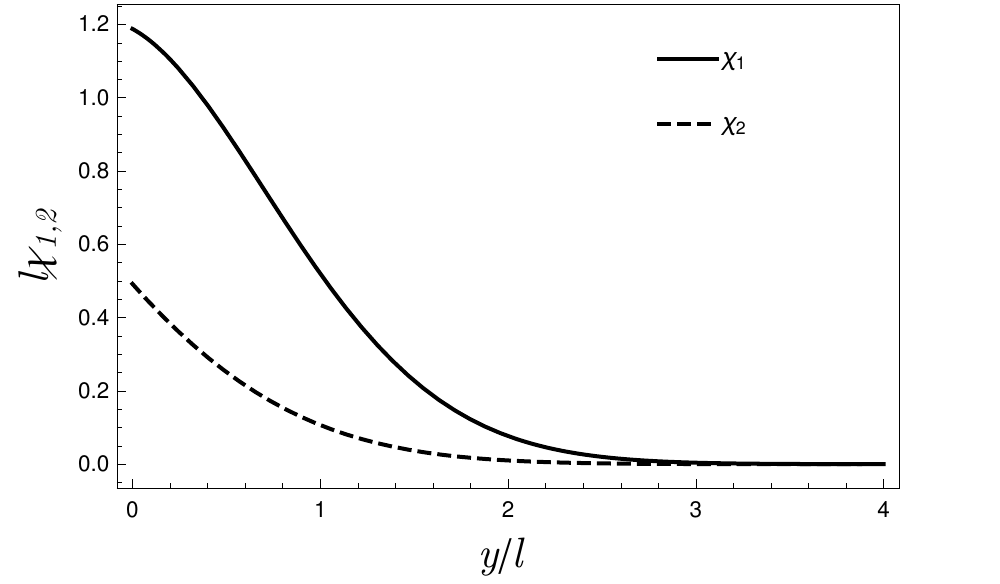} & \includegraphics[scale=0.5]{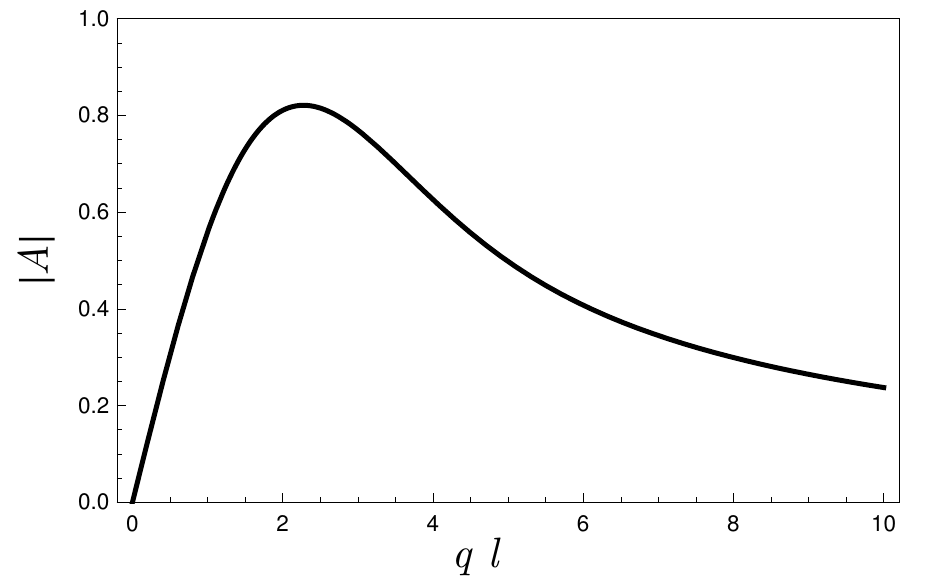}\\
\includegraphics[scale=0.5]{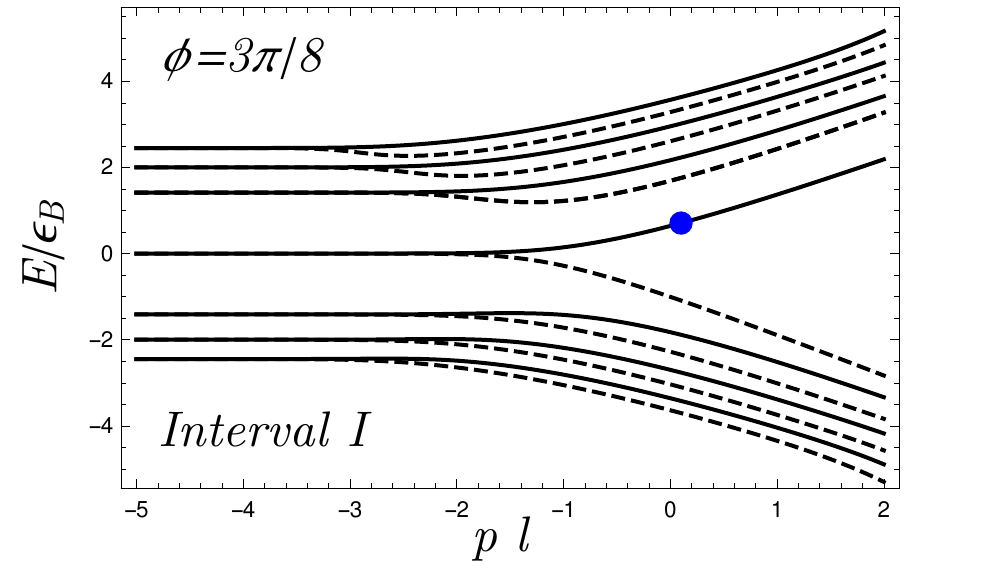} & \includegraphics[scale=0.5]{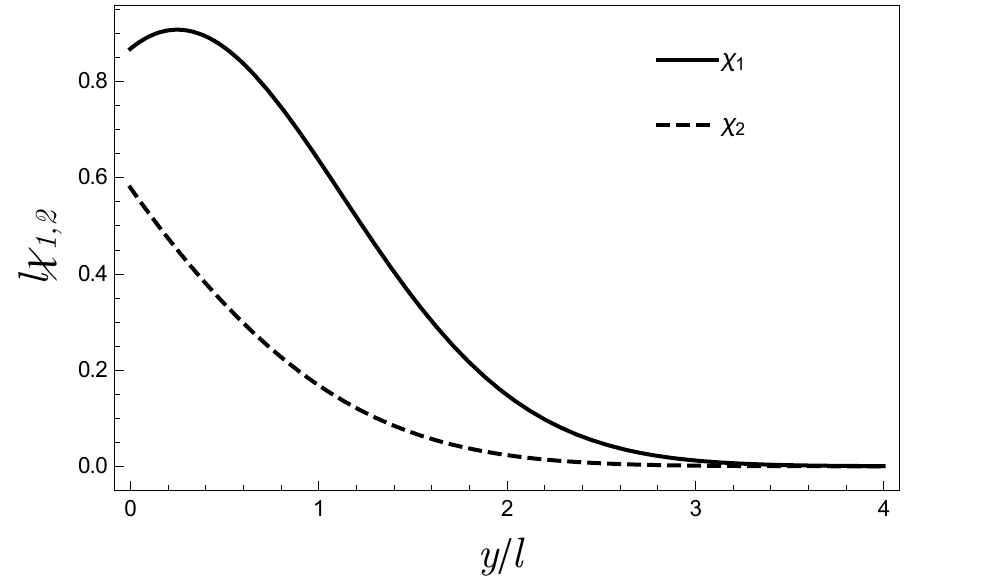} & \includegraphics[scale=0.5]{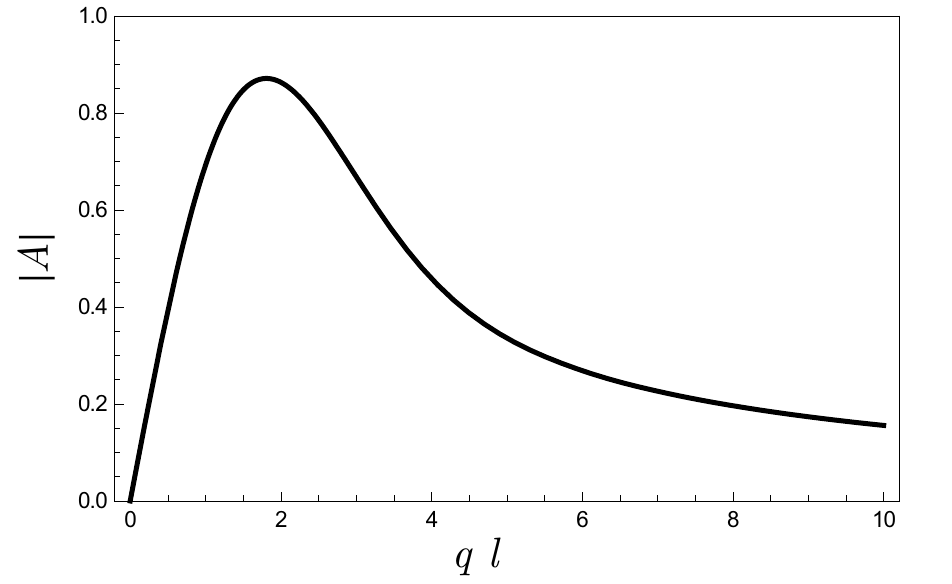}\\
\includegraphics[scale=0.5]{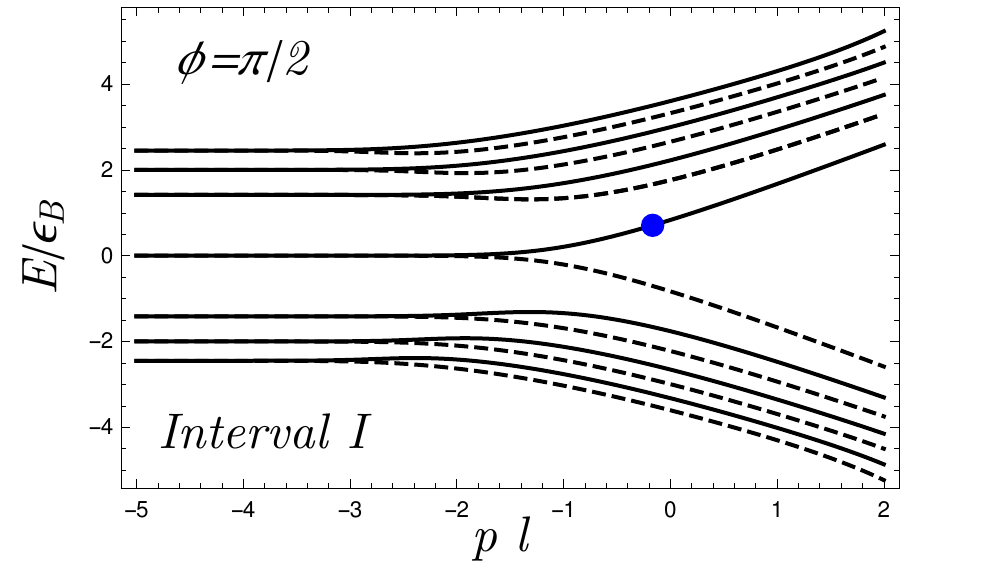} & \includegraphics[scale=0.5]{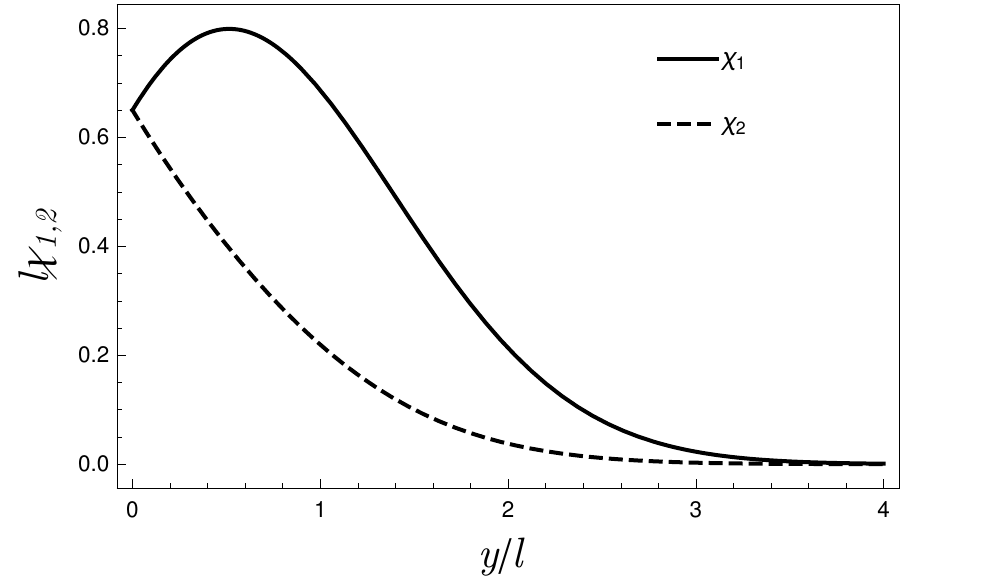} & \includegraphics[scale=0.5]{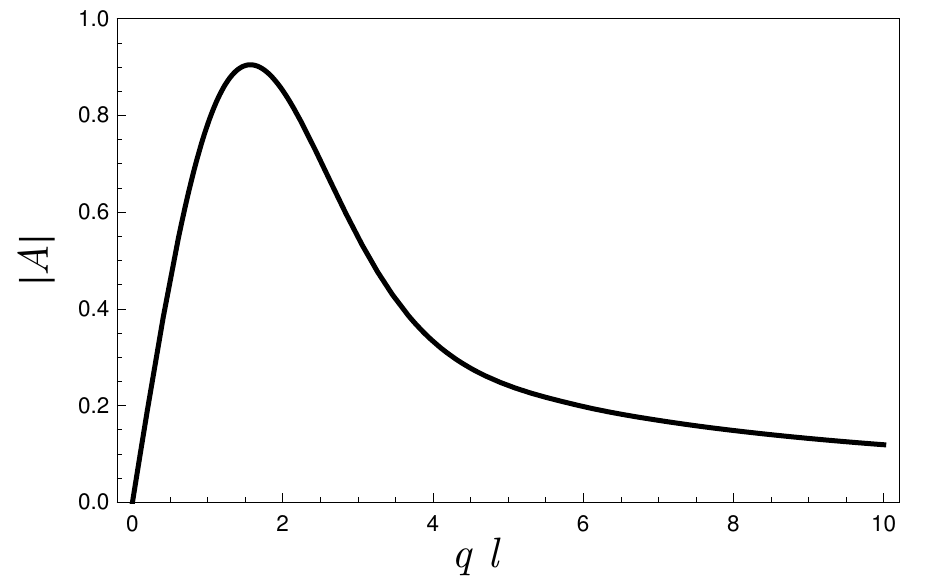}
\ea
$
\caption{\label{fig:FullFigures} 
Left: spectrum of edge states for  $\phi\ge 0$  with $\nu=2$ edge state marked with a dot (solid and dashed lines correspond to $\chi_+$ and $\chi_-$ `valleys'); 
Middle: two components of the wave-functions, $\chi_{1,2}$, for edge states for the bulk filling factor $\nu=+2$  ($E = \eps_B/\sqrt{2} = \frac{\hbar v}{\sqrt{2}l}$).
These plots are also valid for  $\phi \to - \phi$, $\nu = - 2$;
Interchanging $\chi_+ \leftrightarrow \chi_-$  gives the plots for $\phi \to \phi + \pi $ (with the same wave-function components); 
Right: Phonon form-factors for 
the fastest edge state (marked with blue dot) at $E = \eps_B/\sqrt{2}$ .
}
\end{figure*}

\begin{figure*}
$\ba{ccc}
\includegraphics[scale=0.5]{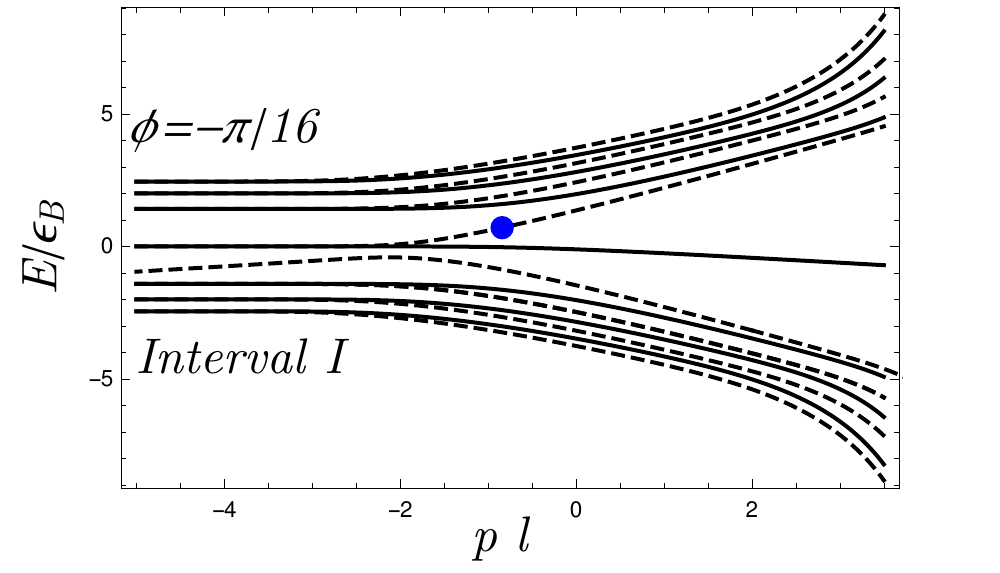} & \includegraphics[scale=0.5]{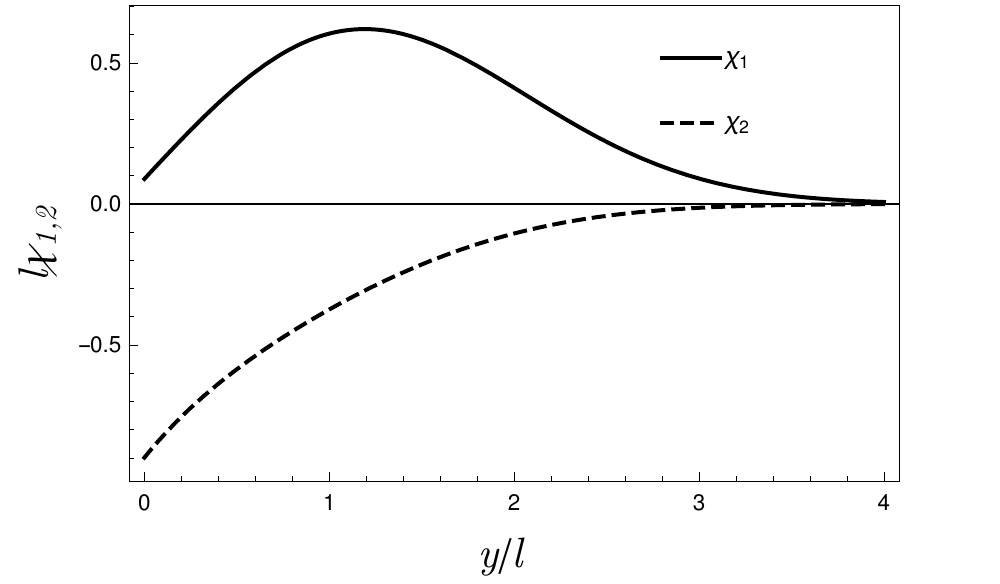} & \includegraphics[scale=0.5]{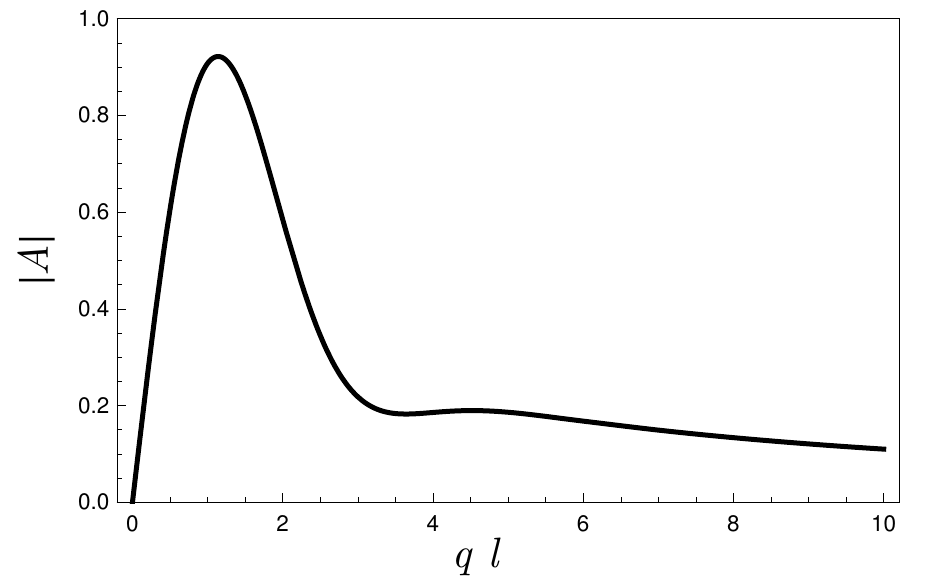}\\
\includegraphics[scale=0.5]{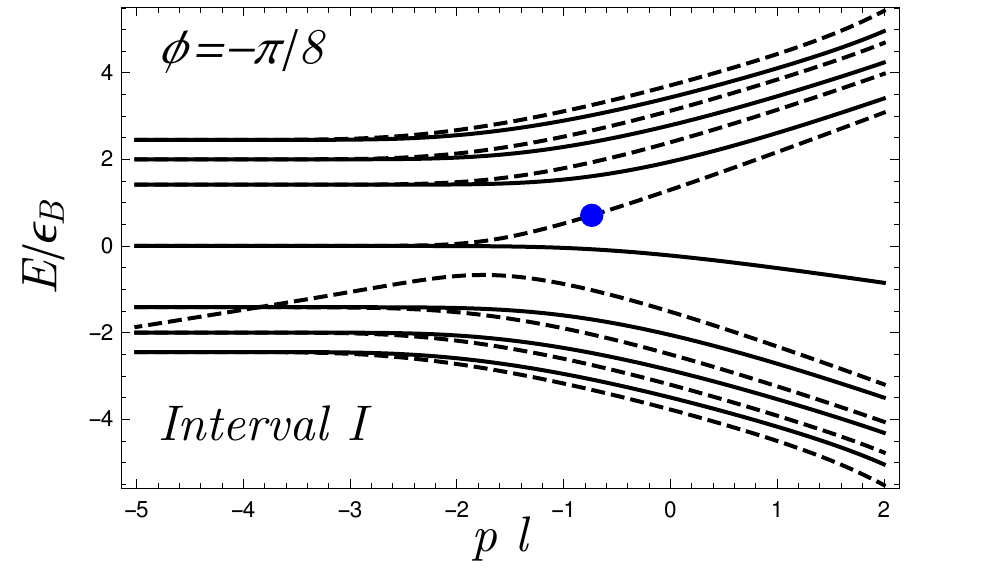} & \includegraphics[scale=0.5]{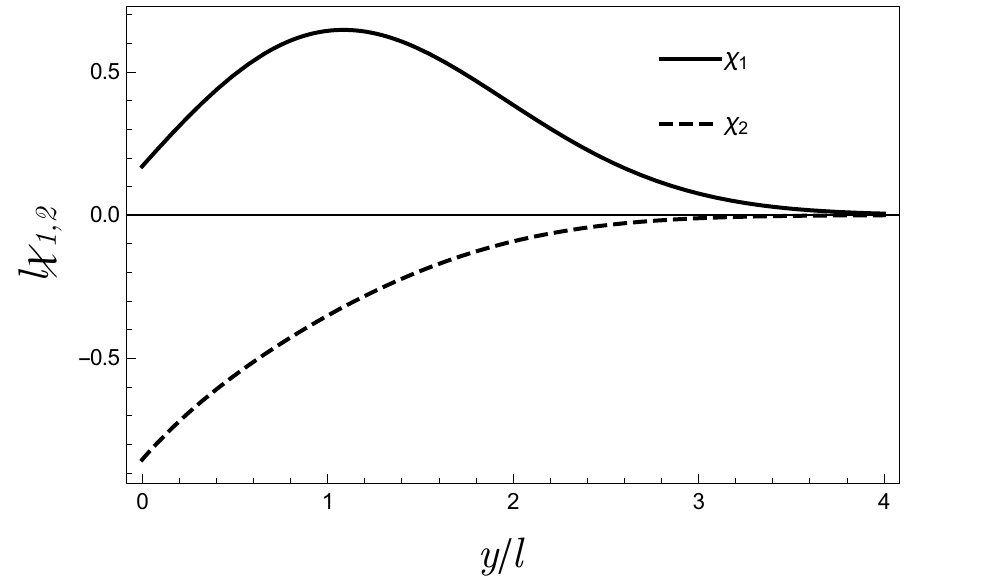} & \includegraphics[scale=0.5]{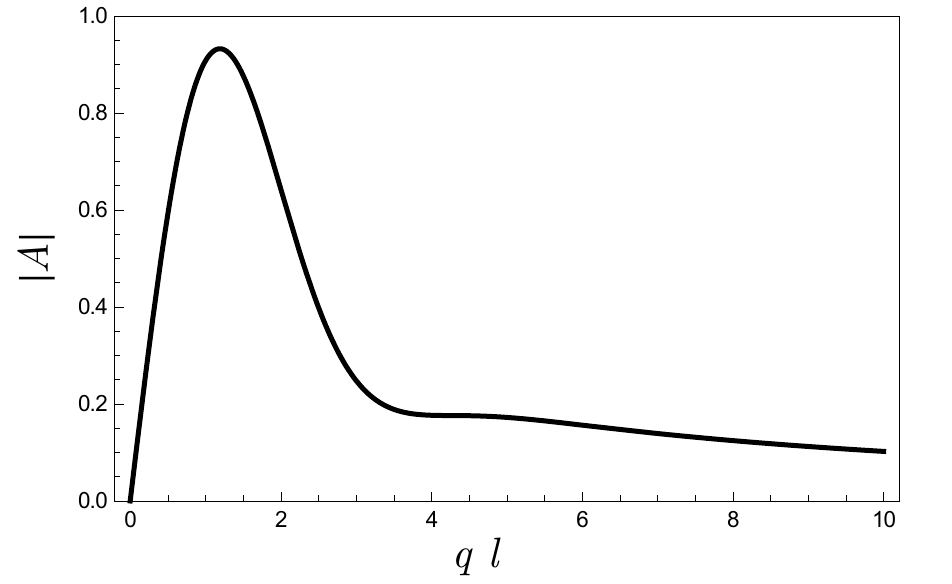}\\
\includegraphics[scale=0.5]{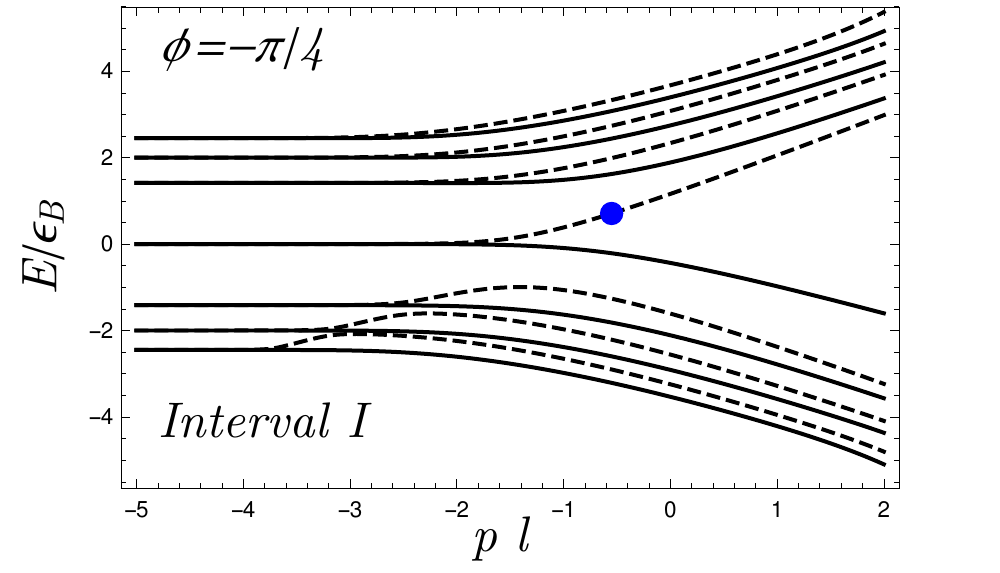} & \includegraphics[scale=0.5]{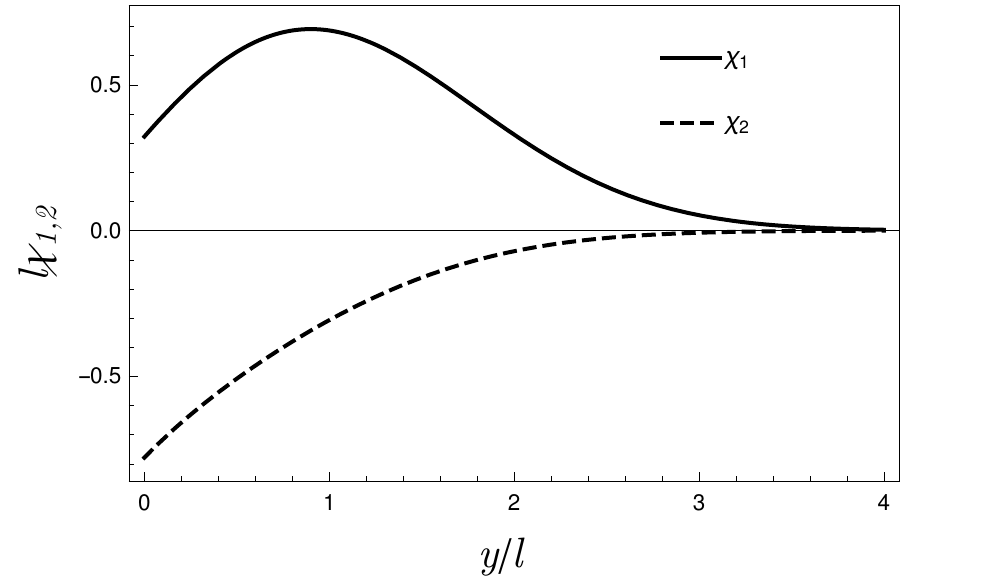} & \includegraphics[scale=0.5]{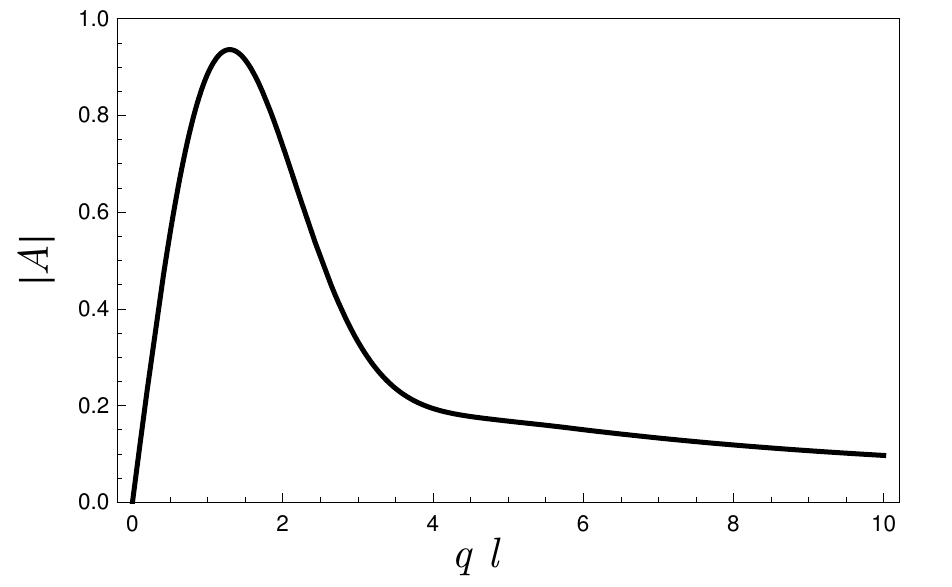}\\
\includegraphics[scale=0.5]{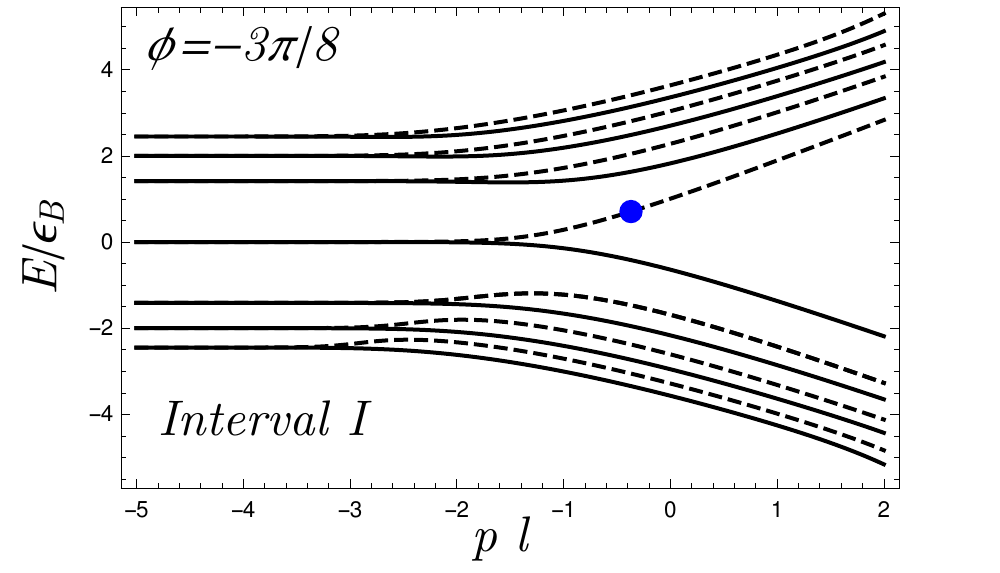} & \includegraphics[scale=0.5]{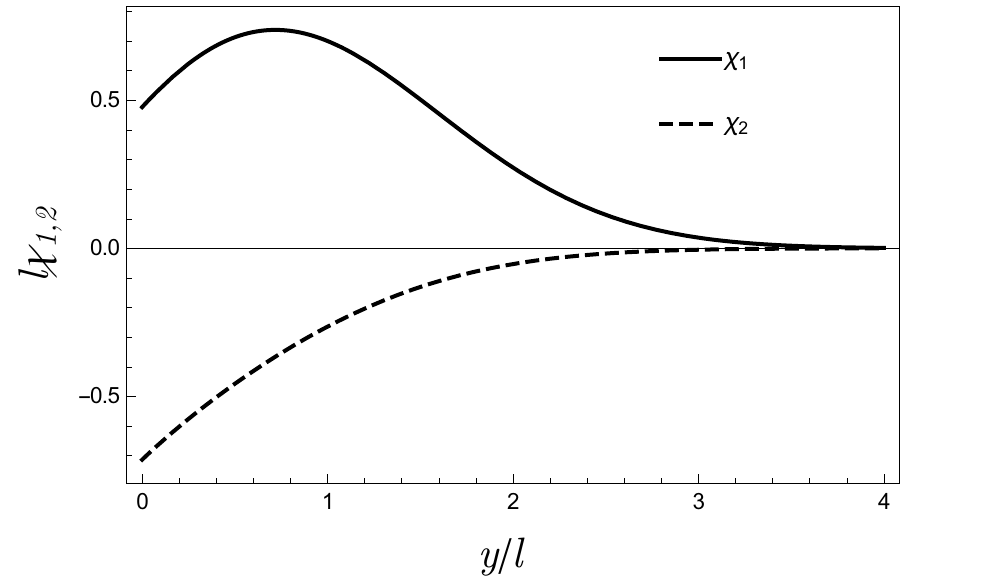} & \includegraphics[scale=0.5]{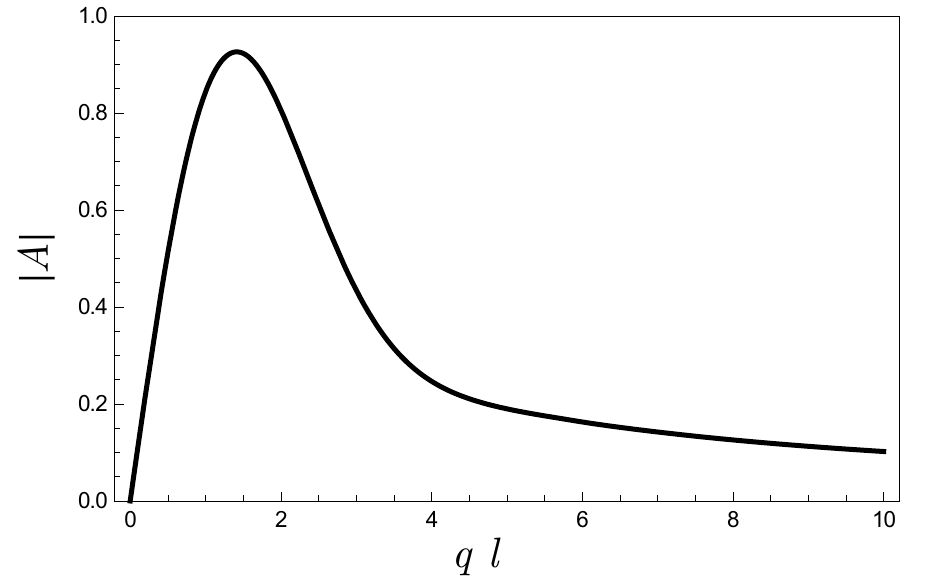}\\
\ea
$
\caption{\label{fig:FullFigures1} 
Left: spectrum of edge states for $\phi<0$ with $\nu=2$ edge state marked with a dot (solid and dashed lines correspond to $\chi_+$ and $\chi_-$ `valleys'); 
Middle: two components, $\chi_{1,2}$, of the wave-functions for edge states for the bulk filling factor $\nu=2$  ($E = \eps_B/\sqrt{2} =  \frac{\hbar v}{\sqrt{2}l}$);
These plots are also valid for $-\phi \to + \phi$ and $\nu = - 2$;
Interchanging $\chi_+ \leftrightarrow \chi_-$  gives the plots for $\phi \to \phi + \pi $; 
Right: Phonon form-factors for 
the $\nu = 2$ edge state (marked with blue dot) at $E = \eps_B/\sqrt{2}$ .
}
\end{figure*}

The spectrum of edge states has been calculated numerically from \eq{BoundaryCond} with the shooting method (tuning of energy $E$ to make the wave-function vanish far from the edge, 
while smoothly evolving in $p$) 
and the corresponding phonon form-factors, $A$, defined in \eq{FormFactorDef},
were computed, see Figs.\ref{fig:FullFigures}, \ref{fig:FullFigures1}.  The resulting wave-function profiles and form factors are shown in Figs.\ref{fig:FullFigures}, \ref{fig:FullFigures1}, and
the computed values of $v_e$ and parameters of asymptotic behavior are shown in Table I.

\section{S2. Edge state thermalization \label{sec:Thermalization}} 
If the phonon cooling of edge electrons were faster then their thermalization, we would have to study the Boltzmann equation to trace the evolution
of electron distribution $n(p)$. The typical e-ph transition goes from the 
tail of hot electrons at $p'$ to the center of the distribution at $p \approx p_F$, with a probability $P_{p' \to p} \sim n(p) |p| |A(p v_e/s)|^2 $.  
From \eq{Heph} we deduce  that for $T_e\gg T_*$, 
$\d_t n(p') = P_{p' \to p} \sim n(p)/|p-p'|$, which leads to a weaker cooling in the high energy tail.
At $T_e \ll T_*$,    $\d_t n(p') = P_{p' \to p} \sim n(p) |p-p'|^3$ and the tails of electron distribution cool faster.  
If we approximate $A(q) \approx const$, which holds in the intermediate temperature range, the phonon cooling  approximately preserves 
the Fermi distribution, which is confirmed by  numerical solution for the temperature range 10 - 100 K. 

\section{S3. Clamped edge of graphene \label{sec:Clamped}}
\newsavebox{\mybox}
\begin{lrbox}{\mybox}
$ 
\tilde A(q) \approx 
\protect\left\{ \protect\ba{cc} -\frac{r_1}{(ql)^2}, & q\gg l^{-1}  \\
                        \sqrt{2}, & q\ll l^{-1}           \protect \ea \protect\right. ,   
$
\end{lrbox}
\newsavebox{\myboxx}
\begin{lrbox}{\myboxx}
$
\ba{c}
f_{\tilde A}\left(\frac{T_e}{T_*},\frac{T}{T_*}=0\right) \approx  \left \{ \ba{cc}  r_2 \, \frac{T_e}{T_*} ,  & T_e \gg T_*; \\
\frac{\pi^2}{30} \frac{T_e^4}{T_*^4}     , & T_e \ll T_*, \ea \right. 
\ea
$
\end{lrbox}
\newsavebox{\myboxxx}
\begin{lrbox}{\myboxxx}
$ 
  T_e(x) \approx \left \{ \ba{cc} T_0  - \frac{12 r_2}{r_0^2}\, T_* \gamma x , & T > T_*; \\
  \frac{T_*}{\sqrt{12 \gamma x - \frac{r_0^2}{r_2}\! \max\left [0,\frac{T_0}{T_*}-1\right] + \max\left[\(\frac{T_*}{T_0}\right)^2,1\right]}} , & T < T_*, \ea \right.  
$
\end{lrbox}
\newsavebox{\myboxff}
\begin{lrbox}{\myboxff}
$\tilde A(q_y) = \int_0^\infty dy \sqrt{2} \cos(q y)  |\psi(y)|^2 $
\end{lrbox}

For the clamped edge of graphene ($u(y=0)=0$), the phonon form-factor is 
\usebox{\myboxff}. 
For its asymptotic behavior  we find
$$\usebox{\mybox}$$ {\red which is weaker / stronger  than free edge case for large / small $q$}.   This leads to
$$\usebox{\myboxx}$$
with the values of $r_2$ given in Table I of SM.
Here, we point out that in graphene with a clamped edge, edge-state electrons are better coupled to lower-energy phonons, 
with $q \lesssim l^{-1}$, whereas in structures with
mechanically free edge  maximum coupling appears at $q \gtrsim l^{-1}$, which promotes cooling at high electron temperatures. 
The resulting  profiles $T_e(x)$  take the form:
$$\usebox{\myboxxx}$$
giving the inverse square root decay of temperature at low temperatures and a linear decay at high temperatures.

\end{document}